%% file: main.tex
\DocumentMetadata{%
	lang=en-US,
	pdfstandard	= A-3u,
	pdfversion	= 1.7,
}

\documentclass[balance,nocopyright,upint,varvw,hyphenate,nolists]{asmejour}

\allowdisplaybreaks 

\usepackage{orcidlink} 
\usepackage{placeins}  
\usepackage{tabularx}  
\input{custom-formatting.tex} 
\usepackage{enumitem}
\usepackage{adjustbox}
\usepackage{multicol}
\usepackage{caption}
\usepackage{pdfpages}

\hypersetup{%
	pdfauthor={Alejandro Pradas G\'omez, Arindam Brahma, Ola Isaksson},
	pdftitle={DUCTILE:\\Agentic LLM Orchestration of Engineering Analysis in Product Development Design Practice},
	pdfkeywords={LLM agents, design automation, engineering practice, aerospace, human-AI collaboration, agentic orchestration, product development},
	pdfsubject={LLM-based agents for aerospace structural analysis within engineering practice environments},
}

\JourName{Mechanical Design}

\begin{document}


\SetAuthorBlock{Alejandro \\Pradas-G\'omez \CorrespondingAuthor\orcidlink{0000-0001-7411-2732}}{%
	Product Development Division,\\
	Department of Mechanical Engineering,\\
	Chalmers University of Technology,\\
	Gothenburg, Sweden\\
	e-mail: alejandro.pradas@chalmers.se\\[6pt]
	GKN Aerospace Engine Systems,\\
	Trollh\"attan, Sweden
}

\SetAuthorBlock{Arindam Brahma \orcidlink{0000-0003-0365-0394}}{%
	Product Development Division,\\
	Department of Mechanical Engineering,\\
	Chalmers University of Technology,\\
	Gothenburg, Sweden\\
	e-mail: arindam.brahma@chalmers.se
}

\SetAuthorBlock{Ola Isaksson \orcidlink{0000-0003-0373-3720}}{%
	Product Development Division,\\
	Department of Mechanical Engineering,\\
	Chalmers University of Technology,\\
	Gothenburg, Sweden\\
	e-mail: ola.isaksson@chalmers.se
}

\title{DUCTILE:\\Agentic LLM Orchestration of Engineering Analysis in Product Development Design Practice}


\begin{abstract}
Engineering analysis automation in product development relies on rigid interfaces between tools, data formats and documented processes. When these interfaces change, as they routinely do as the product evolves in the engineering ecosystem, the automation support breaks. This paper presents a DUCTILE (Delegated, User-supervised Coordination of Tool- and document-Integrated LLM-Enabled) agentic orchestration, an approach for developing, executing and evaluating LLM-based agentic automation support of engineering analysis tasks. The approach separates adaptive orchestration, performed by the LLM agent, from deterministic execution, performed by verified engineering tools. The agent interprets documented design practices, inspects input data and adapts the processing path, while the engineer supervises and exercises final judgment. DUCTILE is demonstrated on an industrial structural analysis task at an aerospace manufacturer, where the agent handled input deviations in format, units, naming conventions and methodology that would break traditional scripted pipelines. Evaluation against expert-defined acceptance criteria and deployment with practicing engineers confirm that the approach produces correct, methodologically compliant results across 10 repeated independent runs. The paper discusses the paradigm shift and the practical consequences of adopting agentic automation, including unintended effects on the nature of engineering work when removing mundane tasks and creating an exhausting supervisory role.
\end{abstract}

\date{\today}

\maketitle 


\input{01-introduction}
\input{02-background}

\input{03-llm-foundations}
\input{04-DUCTILE}

\input{05-case-example}

\input{06-discussion}

\input{07-conclusion}

\section*{Acknowledgment} 

The authors thank the engineers at GKN Aerospace who participated in the experimental study, and Najeem Muhammed for his valuable technical knowledge. The authors acknowledge the support from Faez Ahmed and value the fruitful discussions with all DeCoDE lab members at MIT.

\section*{Funding Data}

The first author's salary was funded by GKN Aerospace, where he is a permanent employee. The Barbro Osher Pro Suecia Foundation (Grant No.\ 90401181) provided financial support for the first author's research visit at MIT.

\section*{Conflict of Interest}

The authors declare no conflicts of interest.

\section*{Data Availability Statement}

The agentic application implementation, evaluation results, and supplemental material, including full session transcripts, are openly available at \url{https://github.com/alex-pradas/DUCTILE} and permanently archived~\cite{pradasgomezDUCTILE2026}.


\bibliographystyle{asmejour}   

\bibliography{bibliography}




\onecolumn
\setcounter{page}{1}
\renewcommand{\thepage}{S-\arabic{page}}
\setcounter{figure}{0}
\renewcommand{\thefigure}{S-\arabic{figure}}
\setcounter{table}{0}
\renewcommand{\thetable}{S-\arabic{table}}
\setcounter{lstlisting}{0}
\renewcommand{\thelstlisting}{S-\arabic{lstlisting}}
\setcounter{section}{0}
\renewcommand{\thesection}{S-\arabic{section}}

\fancyhf{}
\fancyfoot[L]{\large\sffamily\bfseries Supplemental Material}
\fancyfoot[R]{\large\sffamily\bfseries\thepage}

\makeatletter
\renewcommand\section{\@startsection{section}{1}{\z@}%
  {-3.5ex plus -1ex minus -.2ex}%
  {2.3ex plus .2ex}%
  {\normalfont\LARGE\bfseries\sffamily}}
\makeatother

\definecolor{chLeaves50}{HTML}{F5F9EE}
\definecolor{chLeaves700}{HTML}{3F6E1E}
\definecolor{chBrick50}{HTML}{FEF2EE}
\definecolor{chBrick700}{HTML}{912D0F}
\definecolor{chSky50}{HTML}{EEF6FC}
\definecolor{chSky700}{HTML}{196690}
\definecolor{chGranite50}{HTML}{F2F4F5}
\definecolor{chGranite300}{HTML}{8E9DA7}
\definecolor{chCopper50}{HTML}{E6F7F6}
\definecolor{chCopper700}{HTML}{007A71}
\definecolor{chEnergy50}{HTML}{FFF9E6}
\definecolor{chEnergy400}{HTML}{C8A004}
\colorlet{chLeaves10}{chLeaves50!25!white}
\colorlet{chBrick10}{chBrick50!25!white}
\colorlet{chEnergy10}{chEnergy50!25!white}
\colorlet{chGranite10}{chGranite50!25!white}

\newcommand{\userbubble}[1]{%
  \medskip\noindent\null\hfill
  \begin{tikzpicture}[baseline=(current bounding box.north)]
    \node[fill=chLeaves10, draw=chLeaves700, rounded corners=3mm, line width=0.4pt,
          inner sep=5pt, outer sep=0pt]
      {\begin{minipage}{\dimexpr0.92\columnwidth-12pt\relax}\small #1\end{minipage}};
  \end{tikzpicture}\par
}
\newcommand{\agentbubble}[1]{%
  \medskip\noindent
  \begin{tikzpicture}[baseline=(current bounding box.north)]
    \node[fill=chBrick10, draw=chBrick700, rounded corners=3mm, line width=0.4pt,
          inner sep=5pt, outer sep=0pt]
      {\begin{minipage}{\dimexpr0.92\columnwidth-12pt\relax}\small #1\end{minipage}};
  \end{tikzpicture}\par
}
\newcommand{\toolbubbleSM}[1]{%
  \medskip\noindent
  \begin{tikzpicture}[baseline=(current bounding box.north)]
    \node[fill=chEnergy10, draw=chEnergy400, rounded corners=3mm, line width=0.4pt,
          inner sep=5pt, outer sep=0pt]
      {\begin{minipage}{\dimexpr0.92\columnwidth-12pt\relax}\small\ttfamily #1\end{minipage}};
  \end{tikzpicture}\par
}
\newcommand{\responsebubble}[1]{%
  \medskip\noindent
  \begin{tikzpicture}[baseline=(current bounding box.north)]
    \node[fill=chGranite10, draw=chGranite300, rounded corners=3mm, line width=0.4pt,
          inner sep=5pt, outer sep=0pt]
      {\begin{minipage}{\dimexpr0.92\columnwidth-12pt\relax}\footnotesize\ttfamily #1\end{minipage}};
  \end{tikzpicture}\par
}
\newcommand{\sysstatus}[1]{%
  \medskip\begin{center}\small\textit{\textcolor{chGranite300}{#1}}\end{center}
}

\newcommand{\execbubble}[1]{%
  \medskip\noindent\null\hfill
  \begin{tikzpicture}[baseline=(current bounding box.north)]
    \node[fill=chLeaves10, draw=chLeaves700, rounded corners=3mm, line width=0.4pt,
          inner sep=5pt, outer sep=0pt]
      {\begin{minipage}{\dimexpr0.92\columnwidth-12pt\relax}\footnotesize\ttfamily #1\end{minipage}};
  \end{tikzpicture}\par
}

  {-3.5ex plus -1ex minus -.2ex}%
  {2.3ex plus .2ex}%


\begin{center}{\huge\bfseries\sffamily SUPPLEMENTAL MATERIAL}\end{center}\medskip

\section{Repository and Archived Data}
\label{sec:repository}

All source code, configuration files, evaluation scripts, and session transcripts are available in the DUCTILE repository:

\begin{itemize}
    \item \textbf{GitHub}: \url{https://github.com/alex-pradas/DUCTILE}
    \item \textbf{Zenodo archive}: \href{https://doi.org/10.5281/zenodo.18836517}{10.5281/zenodo.18836517}
\end{itemize}

GitHub always contains the latest version, which is propagated to Zenodo for long-term academic archival.

\section{IDE Setup for engineers interacting with agentic application}
\label{sec:ide-setup}

Figure~\ref{fig:ide-screenshot} shows the integrated development environment as configured for the engineer sessions. The left panel displays the project file explorer with the working directory structure, the center panel shows the task description document, and the right panel shows the Claude Code agent interface ready to receive instructions.

\noindent
\begin{minipage}{\textwidth}
  \centering
  \includegraphics[width=\textwidth]{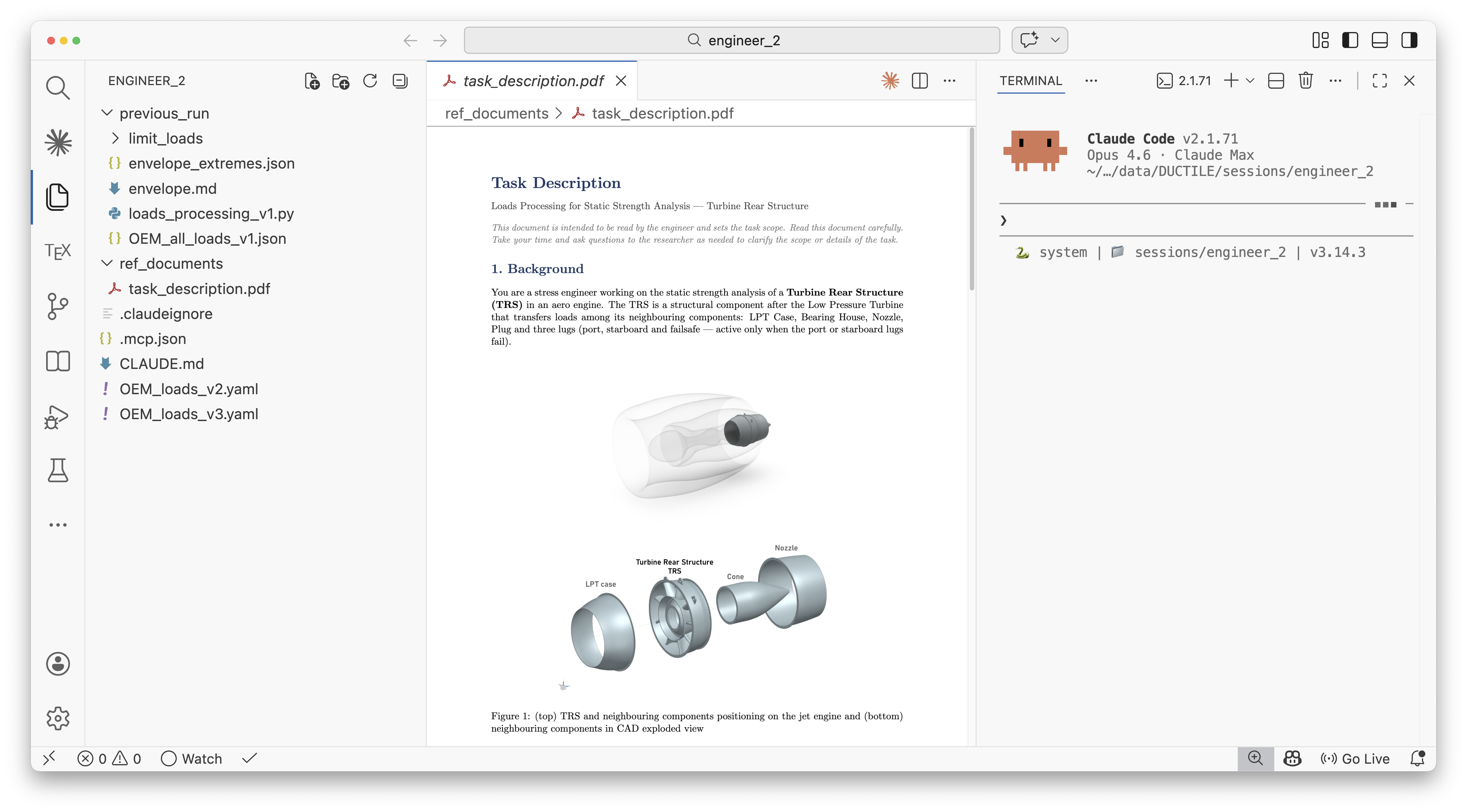}
  \captionof{figure}{Screenshot of Engineer 2 IDE setup for the experiment session}
  \label{fig:ide-screenshot}
\end{minipage}

\clearpage
\section{Agent Configuration}
\label{sec:agent-config}

\bigskip
\noindent
\begin{minipage}{\textwidth}
  \centering
  \includegraphics[width=\textwidth]{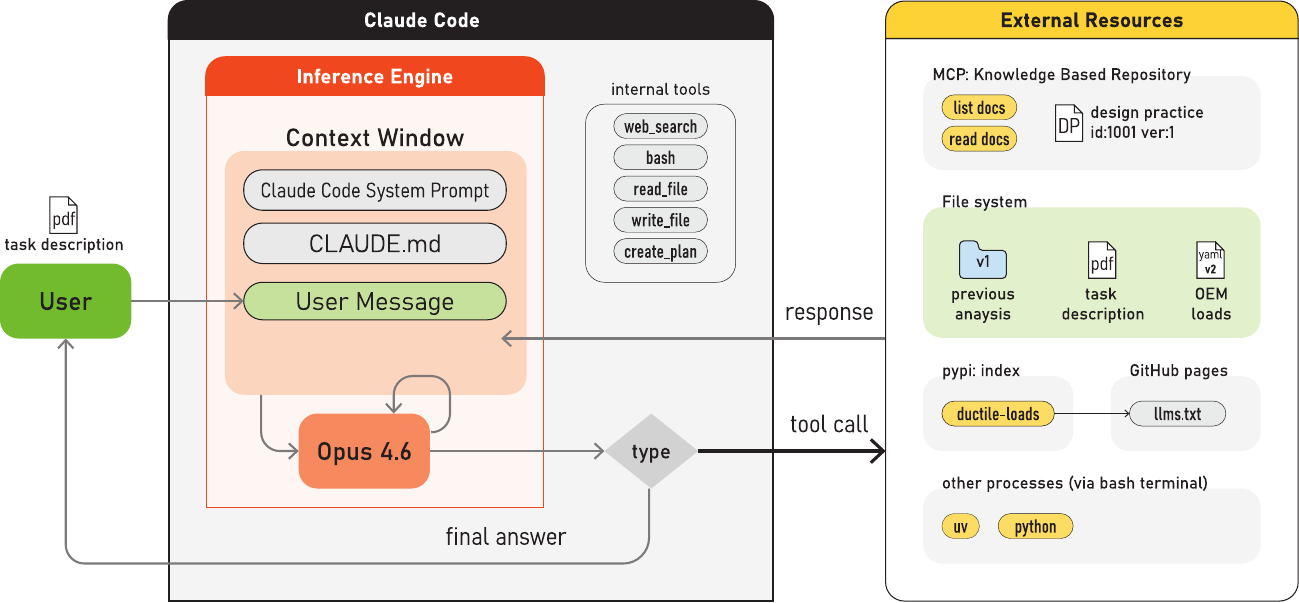}
  \captionof{figure}{Implementation of an agentic structure for the use case. Note how the implementation leverages internal claude code tools with in-house tools (yellow) architecture, that minimizes architectural customization. In addition, the user experience is enhanced as only product data (in green) is required to customize the agent.}
  \label{fig:cc-interface}
\end{minipage}

\begin{multicols}{2}
\begin{lstlisting}[
  caption={Full \texttt{CLAUDE.md} system prompt as deployed in the engineer sessions.},
  label={lst:claude-md},
  basicstyle=\fontsize{7.5}{9}\selectfont\ttfamily,
  numbers=left,
  numberstyle=\tiny\color{gray},
  numbersep=5pt,
  breaklines=true,
  breakatwhitespace=false,
  columns=flexible,
  keepspaces=true,
  frame=single,
  xleftmargin=1.5em,
  xrightmargin=0.5em,
  aboveskip=\medskipamount,
  belowskip=\medskipamount,
  language={},
  showstringspaces=false,
]
# CLAUDE.md

## Role

You are an engineering assistant supporting a stress
engineer. Your sole purpose is to help the engineer
complete their task correctly and efficiently.

## Task Scope

The engineer will present the task to perform. It will
be an engineering task. Searching the web for engineering
references, ANSYS documentation, Python package
documentation, or anything directly relevant to solving
the engineer's task is acceptable.

## Methodology Compliance

Search for design practice documents in the MCP tools
wherever mentioned. A design practice defines the
complete processing chain. You must ensure every step is
followed in the correct order. If the engineer skips a
step or proceeds out of sequence, flag it.

## Tool use

Whatever tool you select to do the job, use the latest
version always, unless specified by the user.

## First Interaction

When the conversation starts, ask the engineer whether
they have used AI coding agents before. Adapt your
behaviour based on their answer:

- If they have experience: Briefly confirm you're ready
  to assist, then wait for them to drive the
  conversation.
- If they have not: Welcome them and offer a few example
  prompts to get started. Examples can range from
  questions on how to execute things, about the tools
  documentation, about methodology of the task or the
  execution of part or the totality of the task.

In either case, do not start processing anything until
the engineer asks you to.

Use your built-in capability to answer questions about
Claude Code.

## Working Directory

Only read files within the current working directory. Do
not navigate to parent directories or read files outside
this folder.

## Behavioural Guidelines

- Be proactive: If you notice something that doesn't
  match the design practice, raise it immediately. Do
  not wait for the engineer to ask.
- Be methodical: Work through the design practice steps
  in order. Suggest the next step when the current one
  is complete.
- Be transparent: When performing calculations or
  transformations, show your reasoning so the engineer
  can verify.
- Respect the engineer's expertise: You are an
  assistant, not the lead. Present findings and
  suggestions, but let the engineer make the final
  decisions.
- Track progress: Keep the engineer aware of which steps
  from the design practice have been completed and which
  remain.
\end{lstlisting}
\begin{lstlisting}[
  basicstyle=\fontsize{7.5}{9}\selectfont\ttfamily,
  numbers=left,
  numberstyle=\tiny\color{gray},
  numbersep=5pt,
  breaklines=true,
  breakatwhitespace=false,
  columns=flexible,
  keepspaces=true,
  frame=single,
  xleftmargin=1.5em,
  xrightmargin=0.5em,
  aboveskip=-0.4pt,
  belowskip=\medskipamount,
  language={},
  showstringspaces=false,
  backgroundcolor=\color{chLeaves50},
  firstnumber=last
]
## Task-Specific Configuration

The certified tool for this task is `ductile-loads`. If
needed, fetch the API reference from:

https://alex-pradas.github.io/ductile-loads/llms-full.txt
\end{lstlisting}
\end{multicols}

\clearpage

The agentic application is configured through a system prompt (\texttt{CLAUDE.md}), completely reproduced in Listing~\ref{lst:claude-md} that defines the agent's role, scope, and behavioural guidelines. Note that the only task specific description is included at the end of the file (in green). In real practice, it would not be specified but pointed to the location where all documentation is stored in-house directly in the tool use section. For complete transparency on the content of the prompt, be aware that Claude Code has its own \href{https://github.com/Piebald-AI/claude-code-system-prompts}{system prompt} that we did not deactivate for the experiment.

The agent is aware of the additional MCP server configurations via the file (\texttt{.mcp.json}) that is read automatically and loads into context the tools to provide access to version-controlled design practice documents. The executable tool itself (\texttt{ductile-loads}) is distributed as a Python package via PyPI, modelling the case company distribution through an internal package index. Readers are welcome to explore it by themselves. The demo repository automatically generates the human (html) and machine (llms.txt and .md files) directly from the codebase.

\section{Task Description}
\label{sec:task-description}

The following three-page document was provided to the engineer as a printed handout before the experiment sessions started. It defines the task scope, available resources, and briefing instructions.

\includepdf[pages=-,pagecommand={\thispagestyle{fancy}}]{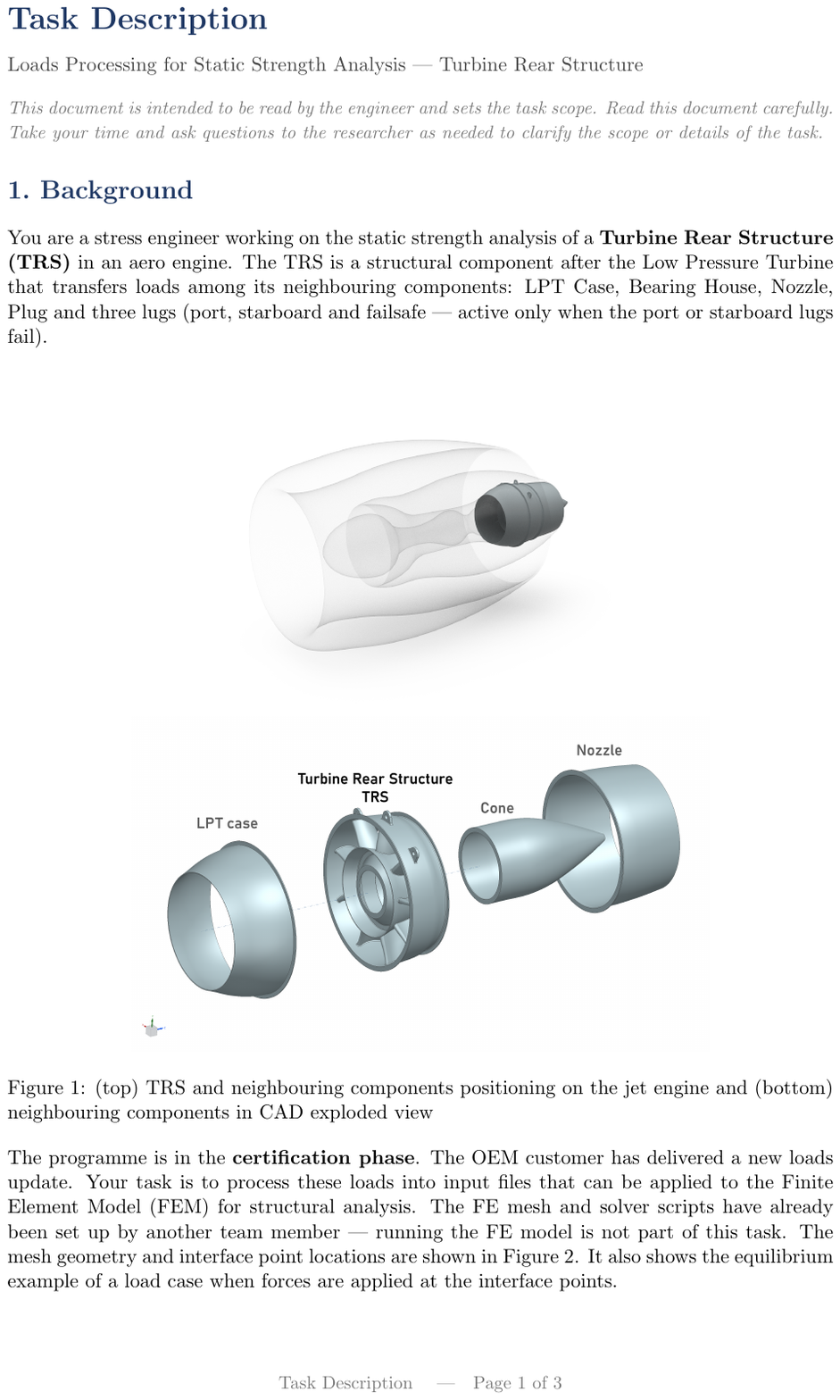}

\section{Engineer 1: Summarized Session Transcript}
\label{sec:engineer1-transcript}

\begin{description}[style=sameline, font=\normalfont]
  \item[\textcolor{chLeaves700}{Green}] Engineer (user) messages
  \item[\textcolor{chBrick700}{Orange}] Agent responses and reasoning
  \item[\textcolor{chEnergy400}{Yellow}] Tool calls and file operations
  \item[\textcolor{chGranite300}{Grey}] Tool/command output
  \item[\textcolor{chCopper700}{Teal}] Engineer terminal execution
\end{description}


\begin{multicols*}{2}
\raggedcolumns

\sysstatus{The engineer spent 10 minutes reviewing the printed task description, the input data files, and the design practice document before starting the session.}

\userbubble{plan this: @ref\_documents/task\_description.pdf}

\agentbubble{Let me read the task description document.}

\toolbubbleSM{Read 1 file (PDF)}

\agentbubble{Let me gather all the context needed to plan this task.}

\toolbubbleSM{Bash(ls) working directory + previous\_run \\[6pt]
ToolSearch: load MCP tools \\[6pt]
design-documents -- get\_document\_content (MCP) \\ (document\_id: 1001, version: 1) \\[6pt]
Read 4 files (v1 script, envelope.md, .inp, YAML)}

\agentbubble{Good, I now have all the context. Here's the plan:

\medskip
\textbf{Task Plan: Loads Processing for TRS Static Strength (v2)}

\medskip
\textbf{Key Issues to Address}
\begin{enumerate}[nosep]
  \item \textbf{Format mismatch} --- OEM delivered YAML; \texttt{ductile-loads} expects JSON.
  \item \textbf{OEM Fx correction} --- Multiply all \texttt{fx} values by \textbf{1.04} before processing.
  \item \textbf{Naming mismatch} --- v2 uses \texttt{lug\_left}/\texttt{lug\_right} instead of \texttt{lug\_port}/\texttt{lug\_starboard}.
  \item \textbf{Units} --- klbs/klbs$\cdot$in $\rightarrow$ N/N$\cdot$m.
  \item \textbf{Bearing exclusion} --- bearing must not appear in output \texttt{.inp} files.
\end{enumerate}

\medskip
\textbf{Processing Chain} (per Design Practice)
\begin{description}[nosep, style=unboxed, leftmargin=1.5em]
  \item[Step 1] Read \& verify OEM delivery --- load YAML, validate. Rename \texttt{lug\_left} $\rightarrow$ \texttt{lug\_port}, \texttt{lug\_right} $\rightarrow$ \texttt{lug\_starboard}.
  \item[Step 2] Apply OEM Fx correction --- multiply all \texttt{fx} $\times$\,1.04.
  \item[Step 3] Verify coordinate system --- engine CS = FEM CS, no transform.
  \item[Step 4] Convert units --- \texttt{convert\_to("N")} --- klbs $\rightarrow$ N.
  \item[Step 5] Envelope downselection --- \texttt{.envelope()}.
  \item[Step 6] Write \texttt{.inp} files --- \texttt{to\_ansys(\ldots, exclude=["bearing"])}.
  \item[Step 7a] Summary envelope table --- \texttt{envelope\_to\_markdown()}.
  \item[Step 7b] Exceedance comparison --- \texttt{compare\_to()} against v1.
\end{description}

\medskip
\textbf{Implementation.}
A single Python script (\texttt{loads\_processing\_v2.py}) using \texttt{uv run -{}-script} with dependencies \texttt{ductile-loads[all]} and \texttt{pyyaml}.

\medskip
Shall I proceed with execution?}

\sysstatus{Crunched for 1 min 39 s}

\userbubble{yes}

\agentbubble{Let me start by fetching the \texttt{ductile-loads} documentation and examining the data more closely, then write the script.}

\toolbubbleSM{WebFetch(https://alex-pradas.github.io/\allowbreak ductile-loads/llms-full.txt) \\[6pt]
Read(previous\_run/envelope\_extremes.json)}

\agentbubble{Now I have everything. Let me write the processing script.}

\toolbubbleSM{Write(loads\_processing\_v2.py) --- 85 lines}

\agentbubble{Now let me run it.}

\toolbubbleSM{Bash(uv run loads\_processing\_v2.py)}

\responsebubble{Delivery: Engine Mount Balanced Loads v2, version 2 \\
Load cases: 100 \\
Renamed lug\_left/lug\_right (200 renames) \\
Applied Fx correction factor 1.04 \\
Units converted: klbs $\rightarrow$ N \\
Envelope downselection: 100 $\rightarrow$ 6 load cases \\
ANSYS .inp files written to limit\_loads/ \\
Exceedance check: new exceeds old = True \\
{[output truncated]}}

\agentbubble{All processing completed successfully. Let me verify the outputs.}

\toolbubbleSM{Bash(ls limit\_loads/) \\[6pt]
Read(limit\_load\_2.inp) \\[6pt]
Bash(ls comparison\_report/) \\[6pt]
Read(v1\_vs\_v2.json)}

\agentbubble{Processing complete. Here's the summary:

\medskip
\textbf{Design Practice steps completed}
\begin{description}[nosep, style=unboxed, leftmargin=1.5em]
  \item[1. Read \& verify] 100 load cases, 7 interfaces. Renamed \texttt{lug\_left} $\rightarrow$ \texttt{lug\_port}, \texttt{lug\_right} $\rightarrow$ \texttt{lug\_starboard}.
  \item[2. Fx correction] All Fx $\times$\,1.04.
  \item[3. Coord system] No transformation needed.
  \item[4. Unit conversion] klbs/klbs$\cdot$in $\rightarrow$ N/N$\cdot$m.
  \item[5. Downselect] 100 $\rightarrow$ 6 load cases (2, 20, 34, 61, 92, 99).
  \item[6. ANSYS export] 6 \texttt{.inp} files in \texttt{limit\_loads/} (bearing excluded).
  \item[7a. Envelope table] \texttt{envelope.md} + \texttt{envelope\_extremes.json}.
  \item[7b. Exceedance] Comparison report + charts in \texttt{comparison\_report/}.
\end{description}

\medskip
\textbf{Exceedance Alert}

New loads exceed the previous v1 envelope. The largest percentage changes:
\begin{itemize}[nosep]
  \item Mz min across all points: $\sim$\textbf{+30\%}
  \item Fx min at bearing/lpt/nozzle/plug: $\sim$\textbf{+15\%}
  \item Fy max/min at lug\_port: \textbf{+11--17\%}
  \item Fy max at lug\_starboard: $\sim$\textbf{+11\%}
  \item Fx max at lug\_starboard: $\sim$\textbf{+14\%}
\end{itemize}

These load increases may impact existing margins of safety and should be flagged to the stress lead.}

\sysstatus{Cogitated for 2 min 53 s}

\sysstatus{The engineer spent 6 minutes reviewing the generated Python script and the output files. After confirming the results were correct, the engineer was satisfied with the task execution.}

\end{multicols*}


\clearpage
\section{Engineer 2: Summarized Session Transcript}
\label{sec:engineer2-transcript}

The transcript below is a faithful representation of the interactions between the engineer and the agent. Message content has been summarized for readability; tool calls and outputs are abbreviated. Full machine-readable transcripts (JSONL) are available in the repository.


\begin{multicols*}{2}
\raggedcolumns

\sysstatus{The engineer spent 8 minutes reviewing the printed task description and the working directory before starting the session.}

\userbubble{Can you create a python file to translate the YAML to a JSON format?}

\agentbubble{Let me read the YAML file to understand its structure.}

\toolbubbleSM{Read(OEM\_loads\_v2.yaml) \\[6pt]
Write(yaml\_to\_json.py) --- 20 lines}

\agentbubble{Created \texttt{yaml\_to\_json.py}. Run it with \texttt{uv run yaml\_to\_json.py}.}

\execbubble{\$ uv run yaml\_to\_json.py}

\responsebubble{Converted OEM\_loads\_v2.yaml $\rightarrow$ OEM\_loads\_v2.json}

\userbubble{ok, now create a file that uses ductile-loads to read the json}

\agentbubble{Let me fetch the \texttt{ductile-loads} API reference first.}

\toolbubbleSM{WebFetch(https://alex-pradas.github.io/\allowbreak ductile-loads/llms-full.txt) \\[6pt]
Write(read\_loads.py) --- 18 lines}

\execbubble{\$ uv run read\_loads.py}

\responsebubble{Name: Engine Mount Balanced Loads v2 \\
Units: klbs, klbs.in | Load cases: 100 \\
{[output truncated]}}

\userbubble{modify the file to convert to newtons, and also rename lug\_right to pilot\_lug\_starboard and lug\_left to pilot\_lug\_port}

\toolbubbleSM{Edit(read\_loads.py) --- rename + convert\_to("N")}

\agentbubble{I used \texttt{rename\_points()} based on the API docs. Let me verify that method exists.}

\userbubble{yes check}

\toolbubbleSM{WebFetch(ductile-loads/llms-full.txt)}

\agentbubble{No \texttt{rename\_points} method exists. I'll rename by iterating through load cases directly.}

\toolbubbleSM{Edit(read\_loads.py) --- iterate + convert\_to("N")}

\execbubble{\$ uv run read\_loads.py}

\responsebubble{Units: N, Nm | pilot\_lug\_port, pilot\_lug\_starboard \\
{[output truncated]}}

\userbubble{ah, I see there are two pilot\_ pilot\ldots\ remove the extra one we added}

\agentbubble{Which rename do you want to remove --- \texttt{pilot\_lug\_port} or \texttt{pilot\_lug\_starboard}?}

\userbubble{just remove the prepend pilot\_}

\toolbubbleSM{Edit(read\_loads.py) --- lug\_left $\rightarrow$ lug\_port, lug\_right $\rightarrow$ lug\_starboard}

\execbubble{\$ uv run read\_loads.py}

\responsebubble{lug\_port, lug\_starboard --- confirmed \\
{[output truncated]}}

\userbubble{how does the envelope in this python package work?}

\agentbubble{The \texttt{envelope()} method reduces your 100 load cases to only the critical ones --- for each point and component it selects the max value (always) and the min value (only if negative). Would you like me to add it?}

\userbubble{yes}

\toolbubbleSM{Edit(read\_loads.py) --- add .envelope() + JSON export}

\userbubble{ok. add now the export to ansys}

\toolbubbleSM{Edit(read\_loads.py) --- add to\_ansys()}

\execbubble{\$ uv run read\_loads.py}

\responsebubble{Envelope: 6 load cases | ANSYS files written to limit\_loads/ \\
{[output truncated]}}

\sysstatus{The researcher verbally reminds the engineer that the agent can access and check the design practice on his own}

\userbubble{ok. I think I am done. Can you check against the design practice 1001 v1?}

\toolbubbleSM{design-documents -- browse\_catalog (MCP) \\[6pt]
design-documents -- get\_document\_content (MCP) \\ (document\_id: 1001, version: 1)}

\agentbubble{Comparing against DP-TRS-LOADS-001 Rev~1:

\medskip
\textbf{Divergences found:}
\begin{enumerate}[nosep]
  \item \textbf{Bearing} --- must be excluded from output.
  \item \textbf{lug\_fairlead} --- should be \texttt{lug\_failsafe}.
  \item \textbf{Exceedance comparison} --- step~7.2 requires comparison against v1.
\end{enumerate}

\medskip
Would you like me to fix these?}

\userbubble{yes fix please}

\toolbubbleSM{Glob(previous\_run/*) \\[6pt]
Read(previous\_run/OEM\_all\_loads\_v1.json) \\[6pt]
Edit(read\_loads.py) --- exclude bearing, rename lug\_failsafe, add compare\_to(v1)}

\userbubble{can you run it?}

\toolbubbleSM{Bash(uv run read\_loads.py)}

\responsebubble{Envelope: 6 cases, 5 points (bearing excluded) \\
WARNING: v2 exceeds v1 envelope \\
\textcolor{red}{ModuleNotFoundError: matplotlib}}

\agentbubble{Needs matplotlib. Let me add the \texttt{[all]} extra to the dependency.}

\toolbubbleSM{Edit(read\_loads.py) --- ductile-loads[all] \\[6pt]
Bash(uv run read\_loads.py)}

\responsebubble{Envelope: 6 cases, 5 points \\
WARNING: v2 exceeds v1 envelope \\
Comparison report generated in comparison\_report/}

\userbubble{ok, I am happy. One final check against the design practice?}

\agentbubble{All processing steps from DP-TRS-LOADS-001 are covered: format conversion, naming, units, envelope, ANSYS export, and exceedance comparison.}

\userbubble{ah, I forgot to multiply fx loads only by 1.04. Can you do that?}

\toolbubbleSM{Edit(read\_loads.py) --- add Fx $\times$ 1.04 before unit conversion}

\execbubble{\$ uv run read\_loads.py}

\responsebubble{Fx correction applied | Envelope: 6 cases \\
WARNING: v2 exceeds v1 envelope \\
{[output truncated]}}

\sysstatus{Engineers checked the output for 3 minutes, and considered the task completed. Total session duration: 22 minutes. The engineer executed scripts 8 times independently, driving the process step by step.}

\end{multicols*}


\end{document}

%% file: custom-formatting.tex
\usepackage{listings}  
\usepackage{tikz} 
\usetikzlibrary{shapes.geometric, arrows.meta, positioning, calc, decorations.pathmorphing}

\definecolor{fcNavy}{HTML}{1F3864}
\definecolor{fcBlueBg}{HTML}{F0F4F8}
\definecolor{fcGreenBorder}{HTML}{4A7C3F}
\definecolor{fcWarnBorder}{HTML}{C67A1F}
\definecolor{fcWarnBg}{HTML}{FFF8F0}
\definecolor{fcGrey44}{HTML}{444444}

\definecolor{chalmersLeaves25}{HTML}{F9FBF5}
\definecolor{chalmersLeaves50}{HTML}{F3F8EC}
\definecolor{chalmersLeaves200}{HTML}{BFD899}
\definecolor{chalmersLeaves700}{HTML}{3F6E1E}  
\definecolor{chalmersGranite50}{HTML}{F2F4F5}
\definecolor{chalmersGranite200}{HTML}{B6C1C8}
\definecolor{chalmersGranite300}{HTML}{8E9DA7}
\definecolor{chalmersBrick50}{HTML}{FEF2EE}    
\definecolor{chalmersBrick400}{HTML}{D4622E}    
\definecolor{chalmersBrick700}{HTML}{B8441A}
\definecolor{chalmersDarkCopper}{HTML}{006C5C}
\definecolor{chalmersDarkCopper50}{HTML}{ECF5F3}  
\definecolor{chalmersSky100}{HTML}{E6F3FB}     
\definecolor{chalmersSky400}{HTML}{468DB6}      
\definecolor{chalmersEnergy50}{HTML}{FFF9E6}    
\definecolor{chalmersEnergy400}{HTML}{C8A004}   
\colorlet{chalmersLeaves10}{chalmersLeaves50!25!white}
\colorlet{chalmersBrick10}{chalmersBrick50!25!white}
\colorlet{chalmersEnergy10}{chalmersEnergy50!25!white}
\colorlet{chalmersGranite10}{chalmersGranite50!25!white}

\newcommand{\chatbubble}[3]{
  \begin{tikzpicture}[baseline=(current bounding box.north)]
    \node[fill=#1, draw=#2, rounded corners=3mm, line width=0.4pt,
          inner sep=5pt, outer sep=0pt] {\begin{minipage}{\dimexpr\linewidth-12pt\relax}#3\end{minipage}};
  \end{tikzpicture}%
}
\newcommand{\thinkbubble}[1]{
  \begin{tikzpicture}[baseline=(current bounding box.north)]
    \node[fill=chalmersGranite10, draw=chalmersGranite200, dashed, rounded corners=1.5mm,
          line width=0.4pt, inner sep=4pt,
          outer sep=0pt, font=\scriptsize\itshape] {\begin{minipage}{\dimexpr\linewidth-16pt\relax}#1\end{minipage}};
  \end{tikzpicture}%
}
\newcommand{\toolbubble}[1]{
  \begin{tikzpicture}[baseline=(current bounding box.north)]
    \node[fill=chalmersEnergy10, draw=chalmersEnergy400, rounded corners=1.5mm,
          line width=0.4pt, inner sep=4pt,
          outer sep=0pt, font=\scriptsize\ttfamily] {\begin{minipage}{\dimexpr\linewidth-16pt\relax}#1\end{minipage}};
  \end{tikzpicture}%
}
\newcommand{\planbubble}[1]{
  \begin{tikzpicture}[baseline=(current bounding box.north)]
    \node[fill=chalmersGranite10, draw=chalmersGranite200, rounded corners=1.5mm,
          line width=0.4pt, inner sep=4pt,
          outer sep=0pt, font=\small] {\begin{minipage}{\dimexpr\linewidth-16pt\relax}#1\end{minipage}};
  \end{tikzpicture}%
}

\newcommand{\leftbubble}[3]{
  \noindent
  \begin{tikzpicture}[baseline=(current bounding box.north)]
    \node[fill=#1, draw=#2, rounded corners=3mm, line width=0.4pt,
          inner sep=5pt, outer sep=0pt] {\begin{minipage}{\dimexpr0.88\columnwidth-10pt\relax}#3\end{minipage}};
  \end{tikzpicture}%
}
\newcommand{\rightbubble}[3]{
  \noindent\null\hfill
  \begin{tikzpicture}[baseline=(current bounding box.north)]
    \node[fill=#1, draw=#2, rounded corners=3mm, line width=0.4pt,
          inner sep=5pt, outer sep=0pt] {\begin{minipage}{\dimexpr0.88\columnwidth-10pt\relax}#3\end{minipage}};
  \end{tikzpicture}%
}
\newcommand{\rightactionbubble}[3]{
  \noindent\null\hfill
  \begin{tikzpicture}[baseline=(current bounding box.north)]
    \node[fill=#1, draw=#2, dashed, rounded corners=3mm, line width=0.8pt,
          inner sep=5pt, outer sep=0pt] {\begin{minipage}{\dimexpr0.88\columnwidth-10pt\relax}#3\end{minipage}};
  \end{tikzpicture}%
}
\newcommand{\toolcall}[1]{
  \begin{tikzpicture}[baseline=(current bounding box.north)]
    \node[fill=chalmersEnergy10, draw=chalmersEnergy400, rounded corners=1.5mm,
          line width=0.4pt, inner sep=3pt,
          outer sep=0pt, font=\scriptsize\ttfamily] {#1};
  \end{tikzpicture}%
}

%% file: 01-introduction.tex
\section{Introduction}
\label{sec:introduction}

Product development relies on an engineering ecosystem with many interacting elements: engineers, data, methods, processes and specialist tools \cite{gerickeCoevolutionToolsMethods2026, Eckert2005, Wynn2021}. In engineering analyses, organizations often try to automate how these interactions occur, for example, through scripts and tool-specific workflows. In practice, designers often repeatedly move between context-specific tools, scripts and methods, adapting the workflow as the requirements, models and interfaces change \cite{Wynn2021, Eppinger1994}. However, such changes, even if minor, can often lead to significant disruptions creating knock-on effects in downstream design activities or changes within the product \cite{Wynn2014, Brahma2022}. For example, because of a tool update, or because of revised requirements or an organizational change. The result is a brittle analysis process \cite{Liu2025}, where significant engineering time is spent on data wrangling \cite{Fernandes2023, Eckert2022} and tool orchestration \cite{Qin2016, Cherukuri2024}, rather than on engineering decisions that directly affect product performance.

In practice, organizations have addressed the brittleness of engineering automation through two main strategies. One is to extend the deterministic automations to add more rules, cover more scenarios, which increases complexity and creates new failure modes as the variation grows \cite{laroccaKnowledgeBasedEngineering2012, verhagenCriticalReviewKnowledgeBased2012a,Vidner2022}. If the automation itself cannot cover the changes, the second approach is to rely on expert engineers to adapt on a per-case basis. This second approach is slow and concentrates critical knowledge in individuals, creating an organizational vulnerability when those individuals are not available \cite{QuintanaAmate2017,ahmedUnderstandingKnowledgeNeeds2004b,Vidner2022}. Neither approach scales well in environments where the workflow evolves alongside the product over a decade-long development process. Approaches integrating machine learning with expert validation have improved knowledge sourcing efficiency in well-structured domains \cite{QuintanaAmate2017}, but remain dependent on predefined data structures, expert-curated feature sets, and rigid methodological pipelines, leaving the underlying brittleness of the workflow unaddressed.

In this paper, a third approach is proposed: Large Language Model (LLM)-based agents as orchestration layers that connect engineers to the verified and trusted domain-specific engineering tools. The agent does not replace the domain-specific software, but instead, interprets the documented design practice, inspects the available data and adapts to the particular context. In doing so, it generates and executes processing code at the engineer request through the same verified and trusted tools that the engineer would use manually. By adding a now standardized and universal orchestration layer on the automation, this approach minimizes the development (and maintenance) burden of tailored workflows and frameworks. With this approach, engineers remain responsible for reviewing the plan, supervising execution and validating outputs as per the latest AI policies and standards of aerospace and automotive certification bodies \cite{EASAArtificialIntelligence2024, ISO/IEC-TR-5469:2024, ISO/PAS-8800:2024}. In this framing, adaptability is provided by the agent, while engineering rigor and traceability are preserved by the deterministic, domain-certified tools.

This paper makes two contributions. First, it presents \mbox{DUCTILE}, an approach for developing, executing and evaluating LLM-based agentic automation of engineering analysis tasks, grounded in the separation of adaptive orchestration from deterministic, verified tool execution~\cite{gomezTeamThreeRole2025}. Second, it provides a test case and evaluates the approach on an industrial load processing task at an aerospace manufacturer, where the agent handled four input deviations that would break traditional automation, across 10 independent runs and two engineers with different supervision styles. The novelty lies in demonstrating that LLM agents can orchestrate real aerospace analysis workflows transparently and compatibly with existing verified tools, while absorbing the input variability that breaks traditional automation. The case, tools and evaluation are provided as open source and free of licenses for other researchers to replicate and improve.

%% file: 02-background.tex
\section{Background}
\label{sec:background}

\subsection{The Tool Adoption Challenge at Aerospace Engineering Companies}
\label{sec:tool-challenge}

Of specific interest in this paper are the challenges of automation in the mechanical engineering analysis, particularly in the aerospace domain \cite{QuintanaAmate2017, Kerley2011}. The engines division of GKN Aerospace (hereafter, the case company) is a Swedish manufacturer of jet engine components and is a risk and revenue-sharing partner for various jet engine Original Equipment Manufacturers (OEMs). The case company routinely performs structural analysis activities to substantiate airworthiness and certification activities, an essential part of the safety-focused industry \cite{AS9100}. These activities follow documented design practices and rely on a combination of commercial solvers, legacy in-house scripts, and an internally developed, evolving ecosystem of modular Python tools \cite{Abollado2017}. The variety of tools in the ecosystem creates a practical challenge that directly relates to the user's experience level. A new engineer, for example, may understand the engineering principles, but may face a steep learning curve in connecting that knowledge to the specific tools and sequences unique to the company's internally established processes. On the other hand, an experienced engineer may know the methodology intimately but may lack the time to explore the documentation and capabilities of a frequently updated tool.

Both these cases highlight a significant and ever-increasing gap between the growing capability of computational tools and the engineering judgment required to use them effectively \cite{KrupaArtScienceCreative}. The engineer's value lies in understanding what the analysis must achieve and whether the results are correctly interpreted. However, in practice, when a design change needs to be evaluated, a significant amount of time is spent on non-engineering tasks such as debugging data format mismatches between modules. 

\subsection{Design Automation in Engineering Analysis}
\label{sec:design-automation}

The challenge described at this case company is not unique to the organization. Engineering design problems are creative, ill-defined and context-dependent, often described as \emph{wicked problems} \cite{rittelDilemmasGeneralTheory1973}. In such problems, the designer often lacks clarity early in the process, and the problem(s) become fully clear only as the solution emerges \cite{crossResearchDesignThinking1992}. Simon \cite{SimonIllDefinedProblems1973} framed design as a process of satisficing under bounded rationality. This definition applies to both the design synthesis and analysis. Vidner et al. \cite{Vidner2022} build upon Ullman's design paradox \cite{ullmanMechanicalDesignProcess2018} to argue that if the design automation is seen as a design activity itself (designing a computational support for the engineer developing a product) then the uncertainty and lack of knowledge on the early phase of the development compounds on the design automation layer.

Gericke et al. \cite{gerickeCoevolutionToolsMethods2026} argue that design methods are not isolated procedures but are embedded in ecosystems of representations, tools, and purposes. These properties have implications for any attempt to automate design work: the automation must accommodate variability, incomplete information, and evolving understanding. Precisely the conditions that rigid, deterministic systems handle poorly. Specific to the aerospace industry, this is a challenge as it creates significant bottlenecks to rapid design iterations, both in the design \cite{baskaranMBSEDrivenMDAOCoArchitecture2026} and manufacturing domains \cite{formentiniConceptualDesignAssembly2022}.

In literature, a number of approaches have been put forward over the years to support engineers in performing analysis of their components, mainly through various forms of computational support for engineering design that have evolved through several generations. We explore three of the most common in aerospace \cite{PradasGomez2023EUCASS}. 

The first category is Knowledge-based engineering (KBE) systems, which encode expert rules and geometric primitives to automate routine design tasks, but they require ontologies \cite{gruberTranslationApproachPortable1993} and design primitives to be defined in advance of the design and analysis activity \cite{laroccaKnowledgeBasedEngineering2012}. In industrial practice, KBE developers and the engineers who perform the analysis are typically not the same people \cite{brimbleMOKAModellingLanguage2000}. The specialized knowledge and effort required to build and maintain KBE applications limit their use to high performing teams in early design concepts, where the design space is sufficiently constrained \cite{tobievandenbergMultidisciplinaryModellingSystem2023}. Further, its use in early phases allows Multi-Disciplinary Optimization (MDO) or design space explorations (DSE) within pre-defined validity ranges, that can be built into models \cite{alHandawiDesignSpaceExploration2024}. However, for the detailed design and certification phases, engineers do not work on pre-determined environments, limits, or design paradigm constraints. The data received from previous processes varies considerably due to other teams generating the data manually, and analysis processes and methods quickly evolve. In these scenarios, KBE approaches are rarely viable \cite{kuglerEvolutionKnowledgebasedEngineering2023a}.

The second category is process integration and design optimization (PIDO) platforms or workflow managers \cite{padulaMultidisciplinaryEnvironmentsHistory2006}. Independent platforms such as modeFRONTIER, ModelCenter, optiSLang, iSight, or CAE-embedded equivalents such as ANSYS Workbench, or equivalents from Siemens, Dassault Systems, etc.\ connect heterogeneous solvers through graphical workflow editors, enabling parametric and MDO studies~\cite{hiriyannaiahInformationManagementCapabilities2008}. The third category is scripted pipelines, which serve a similar purpose of connecting sequences of modular tools. 

Each of these approaches shares a common limitation: the interfaces between components are deterministic and must be specified in advance. For example, a KBE ontology cannot accommodate a design parameter that was not foreseen when the system primitives were built. Or a scripted/visual pipeline produces errors when the OEM delivers loads in a different format. In practice, even rule-based systems with fully explicit logic are described as black boxes \cite{kuglerEvolutionKnowledgebasedEngineering2023a} by the engineers who use them: the complexity of the rule chains either graphical or codified exceeds what a practitioner can trace in complex products and design scenarios, regardless of the system's theoretical transparency \cite{verhagenCriticalReviewKnowledgeBased2012a, kuglerEvolutionKnowledgebasedEngineering2023a}.

\subsection{AI in Engineering Design}

The 2022 JMD Special Issue on Artificial Intelligence and Engineering Design~\cite{allisonSpecialIssueArtificial2022} mapped the intersection of AI and design research along five AI method themes: natural language processing, graph neural networks, generative models, transfer learning and representation learning, and across four design research themes spanning conceptual synthesis, process acceleration, direct estimation and broader process support.

Of the five themes, generative models are the most relevant to the present work, yet they address a fundamentally different problem. Deep Generative Models (DGMs) \cite{regenwetterDeepGenerativeModels2022a} such as GANs, VAEs and diffusion models, produce new design candidates or predict performance fields. For example, \cite{whalenReusableSurrogateModels2022} addresses the DGM as a surrogate model. Computation occurs entirely within the neural network and the output is a design proposal or a surrogate prediction, a narrow problem. In contrast, the central challenge highlighted here focuses on the wide problem that includes the detail phases of product development of safety-critical products. In such cases, specified documented processes must be followed, specific tools must be called in a defined sequence, and inputs and outputs must be traceable through the quality assurance chain \cite{EASAPart21SubpartJ, SAEARP4754B2023}. This orchestration problem falls outside the scope of DGMs or other Machine Learning techniques (e.g. pure numerical regression or classification).

These limitations point toward a class of approaches not captured in earlier AI research applied to engineering design: systems that can mediate between heterogeneous tools, follow documented processes, and adaptively orchestrate analysis workflows. Large language model (LLM)–based agents represent a promising step in this direction. In the previously mentioned JMD special issue~\cite{allisonSpecialIssueArtificial2022}, only one article used a LLM architecture~\cite{yuanLeveragingEndUserData2022} for processing textual product descriptions. 
Basic LLM design synthesis tasks were explored as early as 2024~\cite{pradasgomezLargeLanguageModels2024}, \cite{picardConceptManufacturingEvaluating2025}.
However, a literature review that year on the engineering design journals revealed that most research was around the use of LLMs as tools in the process, not as orchestrators~\cite{gomezTeamThreeRole2025}. Since then the capabilities of the models have evolved. Software engineering took a leading role in developing applications that first autocompleted the sentences, then used tools to read and edit files, and finally developed agentic applications to delegate complete tasks inside the Integrated Development Environment (IDE) or in the Command Line Interface (CLI), such as Claude Code~\cite{ClaudeCode2025}, Codex~\cite{OpenAICodex2025} or OpenClaw~\cite{OpenClaw2025}. 

Recently, Massoudi and Fuge \cite{massoudiAgenticLargeLanguage2026} demonstrated that agentic LLMs can support conceptual systems engineering tasks, including requirements analysis, functional decomposition and architecture trade-offs. Their work shows that LLM agents can reason about design intent when provided with appropriate context. However, their scope is the early design phase, where outputs are conceptual and the integration of in-house tools and methods is not emphasized. 
Mustapha \cite{mustaphaLLMsMechanicsDesignManufacturing2025} also mentions the change in scope of LLMs, from tool to orchestrator, paradigm shift from deterministic to stochasticity, and adaptability or \emph{malleability} - a similar concept to ductility. 

\subsection{Summary: systemic brittleness of existing approaches}\label{sec:backgroundsummary}

A recurring theme across engineering design literature is that methods and tools do not work in isolation, but are embedded within evolving socio-technical systems. As Gericke and Eckert \cite{gerickeCoevolutionToolsMethods2026} note, methods, tools and practices co‑evolve; changes in one element propagate into others. 
Empirical studies on engineering change propagation show that even minor interface shifts can trigger disproportionate rework~\cite{Wynn2014, Eckert2004}. Deterministic automation approaches, including KBE systems, scripted pipelines, and rule‑based workflows, are particularly brittle under such conditions because they rely on predefined assumptions about data structures and process sequences. When these assumptions no longer hold, failures emerge that are difficult for engineers to diagnose, leading many to perceive formally transparent systems as practical ``black boxes''  \cite{verhagenCriticalReviewKnowledgeBased2012a}. The limitations of deterministic systems also echo earlier observations that design tools should support adaptive, intelligent behavior rather than encode rigid procedures \cite{cross2025natural}, yet this has still not been realized in current industrial practice.

The background literature, therefore, indicates a common need for an intermediary that is capable of accommodating variability in tools, data, and practice. The literature further shows that the brittleness of previous approaches arises not only from isolated technical faults but also from the inability of deterministic systems to adapt to the normal variability and evolution of engineering practice. This is a central limitation that the LLM‑based orchestration layer proposed in this paper, seeks to address. The central insight from literature is therefore summarized as a core principle which is used to guide the development of the method:

\medskip
\emph{\textbf{Central Orchestration Principle:} Effective engineering orchestration requires an intermediary capable of accommodating variability in data, tools and practices, rather than relying on rigid, predetermined interfaces.}
\medskip

\begin{table*}[hbt]
\centering
\small
\caption{Requirements for DUCTILE and supporting literature}
\label{tab:ductile-reqs}
\begin{tabular}{p{0.06\linewidth} p{0.25\linewidth} p{0.5\linewidth} p{0.10\linewidth}}
\hline
\textbf{Req.} & \textbf{Requirement name} & \textbf{Description} & \textbf{Supporting literature}\\
\hline
\textbf{R1} & Inspectability &
Agentic application (prompts, plans, tool calls, intermediate artifacts) must be directly inspectable by engineers/auditors; no hidden state. &
\cite{EASAArtificialIntelligence2024,ISO/IEC-TR-5469:2024} \\
\textbf{R2} & Reproducibility &
Model, prompt, tool versions and inputs must be pinned and logged; runs are reproducible or cause‑attributable despite stochastic LLM steps. &
\cite{ISO/IEC-TR-5469:2024,EASAArtificialIntelligence2024} \\
\textbf{R3} & Deterministic execution boundary &
Safety/engineering computations run in verified, deterministic tools; the LLM performs orchestration and code generation only. &
\cite{ISO/IEC-TR-5469:2024,mustaphaLLMsMechanicsDesignManufacturing2025} \\
\textbf{R4} & Traceability \& auditability &
Every artifact is trace‑linked to the generating step (arguments, stdout/stderr, checksums, timestamps) to support do/check/approve reviews. &
\cite{EASAArtificialIntelligence2024,ISO/IEC-TR-5469:2024} \\
\textbf{R5} & Data governance &
Deployment mode (on‑premises/API) and data flows comply with confidentiality/export/privacy rules and are justified. &
\cite{EASAArtificialIntelligence2024,ISO/IEC-TR-5469:2024} \\
\textbf{R6} & Human oversight \& accountability &
Engineers review plans and sign off on outputs; autonomy is limited to orchestration, not engineering judgment. &
\cite{EASAArtificialIntelligence2024} \\
\textbf{R7} & Robustness to routine variability &
Orchestration handles foreseeable variation (formats, units, naming, minor method updates) without modifying certified tools. &
\cite{Wynn2014,verhagenCriticalReviewKnowledgeBased2012a,Eckert2004,caputoAnalyzingRealOptions2022,mustaphaLLMsMechanicsDesignManufacturing2025,elghSupportingManagementMaintenance2008,hjertbergManagingDependenciesHeterogeneous2016} \\
\textbf{R8} & Observability &
The system emits traces for model calls and tool invocations (arguments, latency, tokens/cost) sufficient to diagnose failures. &
\cite{Logfire2024,dongTaxonomyAgentOpsEnabling2024} \\
\textbf{R9} & Minimal coupling &
Tool integration is lightweight/externalized (CLIs/typed APIs); generated code is readable and runnable by engineers. &
\cite{verhagenCriticalReviewKnowledgeBased2012a,laroccaKnowledgeBasedEngineering2012,elghSupportingManagementMaintenance2008,hjertbergManagingDependenciesHeterogeneous2016} \\
\textbf{R10} & Evaluation and pass\textasciicircum{}k &
Reliability is demonstrated on curated cases with deterministic checks and/or LLM‑as‑a‑judge, using repeated independent runs (pass\textasciicircum{}k) proportionate to risk. &
\cite{liuAgentBenchEvaluatingLLMs2025,yaoTauBenchBenchmarkToolAgentUser2025,MMPDS} \\
\textbf{R11} & Change control &
Any change to model, prompts, tools, or design practice triggers re‑evaluation on the same cases before deployment. &
\cite{ISO/IEC-TR-5469:2024,EASAArtificialIntelligence2024} \\
\textbf{R12} & Documentation quality &
Tool/method documentation is sufficiently structured for agents to follow procedures; ambiguities are revised or clarified. &
\cite{hsiehToolDocumentationEnables2023,yuanEASYTOOLEnhancingLLMbased2024} \\
\hline
\end{tabular}
\end{table*}

From the survey of literature, a set of requirements for engineering automation can be extracted; Table \ref{tab:ductile-reqs} summarizes twelve requirements, R1-R12 with brief descriptions and supporting references.

%% file: 03-llm-foundations.tex
\section{LLM Agents: Technical Foundations}
\label{sec:llm-foundations}

Building on the central orchestration principle articulated in Sec. \ref{sec:backgroundsummary}, in this section, we now examine how large language models (LLMs) in agents provide such an adaptive intermediary. So far, we have avoided defining the term Agent. Historically in AI, agents have been defined in many ways, including \emph{tools whose interactions are via indirect management} \cite{kayUserInterfacePersonal1990,wooldridgeIntelligentAgentsTheory1995a, maesAgentsThatReduce1994}. In the context of LLMs, as the technology matures, the term continues to evolve reflecting different characteristics and challenges. Earlier work defined LLM agents as a \emph{[...] task oriented combination of LLM models and the infrastructure to access tools, retrieve context information, and store memory of its interactions.} \cite{pradasgomezEvaluationDifferentLarge2024}.
While every LLM and framework provider has a variation of this definition, this paper uses the term that is gaining traction lately, which involves a description of what it does rather than what it is: \emph{An LLM agent runs tools in a loop to achieve a goal} \cite{willisonThinkAgentMay}. In this paper, we distinguish between the \emph{agentic framework}, the software toolkit that provides the inference loop and tool-calling infrastructure (e.g.\ Claude Code, Pydantic-AI), and the \emph{agentic application}, the configured deployment that combines a specific model, system prompt and tool connections for a given engineering context. Sec.~\ref{sec:DUCTILE} describes how an agentic application is developed, executed, and evaluated.

We focus solely on external, off-the-shelf large language models, recognizing that most engineering organizations, especially medium and small firms lack the resources to develop or fine-tune such models in house.
The context window is central to this capability: since externally developed models have no built-in knowledge of an organization's internal tools and processes, the information available in the context window determines what the model can draw upon when generating each token. 

LLMs offer a flexible inference mechanism that can interpret context, reconcile heterogeneous representations, and mediate between evolving tools and engineering intent \cite{mustaphaLLMsMechanicsDesignManufacturing2025}. Recent advances have further strengthened their suitability for orchestration through two capabilities that have matured rapidly: structured reasoning modes (often referred to as \emph{thinking} or \emph{deliberate reasoning}) and tool calling, which enables controlled interactions with deterministic external software. This subsection provides a brief introduction to these mechanisms for a mechanical engineering audience.

\subsection{Response Generation}

\begin{figure}[t]
    \centering
    \includegraphics[width=\columnwidth]{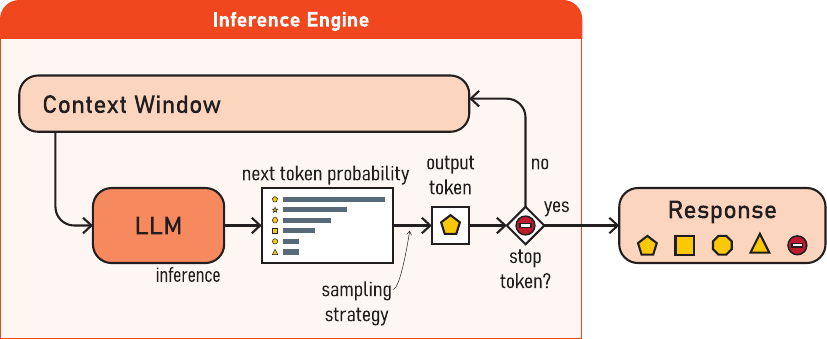}
    \caption{Schematic of the LLM inference loop. The model takes the full context window as input, produces a probability distribution over the next token, samples one token according to a sampling strategy, appends it to the context window, and repeats until a stop token (<EOS>) is generated. The resulting sequence of tokens forms the response.}
    \label{fig:inference-engine}
\end{figure}

A transformer-based LLM \cite{vaswaniAttentionAllYou2017} sequentially generates a token at a time through an autoregressive process, see Fig.~\ref{fig:inference-engine}. At each step, the model takes the entire context window: the system prompt, the user message and all tokens generated so far; and produces a probability distribution over the vocabulary of possible next tokens. A sampling strategy selects one token from this distribution, and the selected token is appended to the context window for the next iteration. This loop continues until a stop token is sampled, at which point the accumulated sequence of tokens is returned as the response. Other recent architectural language models, namely diffusion \cite{labsMercuryUltraFastLanguage2025}, substitute the auto-regressive sequential token generation with a series of refinement steps. Both the architecture and the process are fundamentally non-deterministic: the same input can produce different responses depending on the sampling strategy and its stochastic parameters. 

\subsection{Thinking Mode}

\begin{figure}[t]
\chatbubble{chalmersLeaves50}{chalmersLeaves200}{%
  {\sffamily\bfseries\scriptsize User}\par\smallskip
  \small If a T-shirt takes 5 hours to dry, how long does it take to dry 3 T-shirts?%
}
\vspace{1pt}

\noindent
\begin{minipage}[t]{0.47\columnwidth}
\chatbubble{white}{chalmersGranite200}{%
  {\sffamily\bfseries\scriptsize No thinking}\par\smallskip
  \small Each T-shirt takes 5 hours, so 3 T-shirts take $3 \times 5 = 15$ hours to dry.

  \medskip
  \hfill{\color{chalmersBrick700}\bfseries\sffamily 15 hours \texttimes}%
}
\end{minipage}%
\hfill
\begin{minipage}[t]{0.47\columnwidth}
\chatbubble{white}{chalmersGranite200}{%
  {\sffamily\bfseries\scriptsize Thinking mode}\par\smallskip
  \thinkbubble{$3 \times 5 = 15$ hours to dry\ldots\ Hold on! If the user is hanging them at the same time\ldots\ Wait! the user did not specify... usually laundry is done at the same time so I should assume that the ...}%
  \smallskip

  \small 5 hours. Assuming you have enough space, all three dry simultaneously.

  \medskip
  \hfill{\color{chalmersDarkCopper}\bfseries\sffamily 5 hours \checkmark}%
}
\end{minipage}
\caption{Comparison of LLM responses with and without thinking mode enabled. The thinking scratchpad (dashed box) allows the model to self-critique and reconsider the autoregressive initial response before committing to a final answer.}
\label{fig:thinking-mode}
\end{figure}

The transformer autoregressive nature of token generation introduces a limitation: once the model commits to an initial direction, subsequent tokens are conditioned on that choice, making it difficult to reverse course \cite{zhangHowLanguageModel2024}. This can lead to fixations or hallucinations \cite{jiSurveyHallucinationNatural2023}. Thinking mode is a mitigation strategy developed during model post-training using reinforcement learning \cite{zelikmanStarBootstrappingReasoning2022,snellScalingLLMTestTime2024}, and first experienced by the general public with OpenAI's o1 model in September 2024 \cite{openaiO1SystemCard2024} and subsequent competitor models. 
Practically, the so-called thinking mode provides the model with a scratchpad in which self-critique is rewarded. Specific token sequences allow the model to reconsider its initial direction, explore alternative interpretations and select the most appropriate response before committing to a final answer \cite{weiChainofThoughtPromptingElicits2022}, see Fig.~\ref{fig:thinking-mode}. For engineering orchestration, thinking mode is important because it allows the model to reason at inference time about ambiguous or unexpected situations before acting on them, increasing the probability of a successful final answer \cite{snellScalingLLMTestTime2024}.

\subsection{Tool Calling}

\begin{figure}[t]
\chatbubble{chalmersLeaves50}{chalmersLeaves200}{%
  {\sffamily\bfseries\scriptsize Engineer}\par\smallskip
  \small What is the yield strength of Ti-6Al-4V at 400\,°C?%
}
\vspace{1pt}

\noindent
\begin{minipage}[t]{0.47\columnwidth}
\chatbubble{white}{chalmersGranite200}{%
  {\sffamily\bfseries\scriptsize No tool calling}\par\smallskip
  \small Ti-6Al-4V has a yield strength of approximately 880\,MPa.

  \medskip
  \hfill{\color{chalmersBrick700}\bfseries\sffamily 880\,MPa \texttimes}%
}
\end{minipage}%
\hfill
\begin{minipage}[t]{0.47\columnwidth}
\chatbubble{white}{chalmersGranite200}{%
  {\sffamily\bfseries\scriptsize Tool calling}\par\smallskip
  \toolbubble{materials\_db("Ti-6Al-4V", T=400, units=degC)\par\smallskip \textrightarrow\ $\sigma_y = 620$\,MPa}%
  \smallskip

  \small At 400\,°C, Ti-6Al-4V has a yield strength of approximately 620\,MPa.

  \medskip
  \hfill{\color{chalmersDarkCopper}\bfseries\sffamily 620\,MPa \checkmark}%
}
\end{minipage}
\caption{Comparison of LLM responses with and without tool calling. Without tools, the model relies on parametric knowledge from training weights, returning a room-temperature value. With tool calling, the model queries an external materials database for temperature-dependent data, grounding its response on verified external information.}
\label{fig:tool-calling}
\end{figure}

Tool calling enables the model to ground its responses on external data and computations, beyond what is encoded in its training weights or present in the context window \cite{schickToolformerLanguageModels2023}. Models are trained to recognize when a question requires external information and to generate a structured output (a function name and arguments) instead of a direct answer. The model does not execute the tool itself. In practice, it performs a classification task, selecting which of the available tools is best suited for the current step and inferring the appropriate arguments from the context window. Fig.~\ref{fig:tool-calling} illustrates this with an engineering example.

Wei et al.~\cite{weiAgenticReasoningLarge2026} identify three styles of tool integration in agentic systems: (a) in-context, where the model reads documentation and generates calls zero-shot; (b) post-training, where the tool-use strategies are fine-tuned into model weights, and (c) orchestration-based where a planning layer coordinates multi-tool workflows. The present work uses the first and third; the agent reads tool documentation to learn about unfamiliar APIs and then plans multi-step workflows across multiple tools. Post-training integration is deliberately excluded because engineering tools evolve and must remain independently verified and versioned. Fine-tuning the model to internalize the tool signature or APIs is, as mentioned earlier, possible but unrealistic in practice for most engineering companies.

\subsection{Orchestration}
These three mechanisms, inference, thinking mode, and tool calling, establish a two-layer architecture. The LLM selects actions and adapts to context; the external tools execute deterministically. Yao et al.~\cite{yaoReActSynergizingReasoning2023} formalized this interleaving of reasoning traces and tool actions in the ReAct framework, showing that the model decides at each step whether to reason further, act on the environment, or return a final answer. This pattern has been demonstrated beyond engineering: Shen et al.~\cite{shenHuggingGPTSolvingAI2023} showed that an LLM can act as a controller that plans tasks, selects specialized AI models and orchestrates their execution across vision, speech, and language domains. 

Thinking and tool calling capabilities are pre-requisites for orchestration. Agentic reasoning literature distinguishes two forms of scaling at inference time \cite{weiAgenticReasoningLarge2026}: \emph{scaling test-time computation}, where the model reasons longer within its context window (mapping to thinking) and \emph{scaling test-time interaction}, where the model interacts with external tools and receives feedback from the environment.

\subsection{LLM Agents in Engineering Practice}
\label{sec:llm-engineering}

Developments in the application of LLMs also come from commercial software vendors, who have begun embedding LLM-based assistants within their products. Simulation platforms such as ANSYS \cite{AnsysCopilotWorkFasterSmarter} or COMSOL \cite{COMSOLUsingChatbotWindow} now offer interfaces that support engineer interactivity within the GUI.

These assistants are, however, tightly coupled to their host application. They have access to the model being edited but cannot access external design practice repositories, in-house tools, or data from other stages of the analysis process. This closed-ecosystem constraint limits their applicability to the engineering orchestration problem described in Sec. \ref{sec:tool-challenge}, where the challenge is to bridge across multiple tools and data sources. Emerging service vendors such as COSMON \cite{COSMON} are developing agentic interfaces that connect LLMs to commercial CAD and simulation software. These solutions are interesting for a no-code solution. However, due to the requirements imposed of inspectability (R1), reproducibility (R2) and traceability and auditability (R4) these solutions are discarded as the companies do not have direct access to adapting and therefore adding another potential layer of brittleness.

To address this gap, this paper argues that a minimal and generic agentic application shall be developed at each company.

%% file: 04-DUCTILE.tex
\section{The DUCTILE approach to agentic orchestration}
\label{sec:DUCTILE}

\begin{figure*}[hbt]
    \centering
    \includegraphics[width=\textwidth]{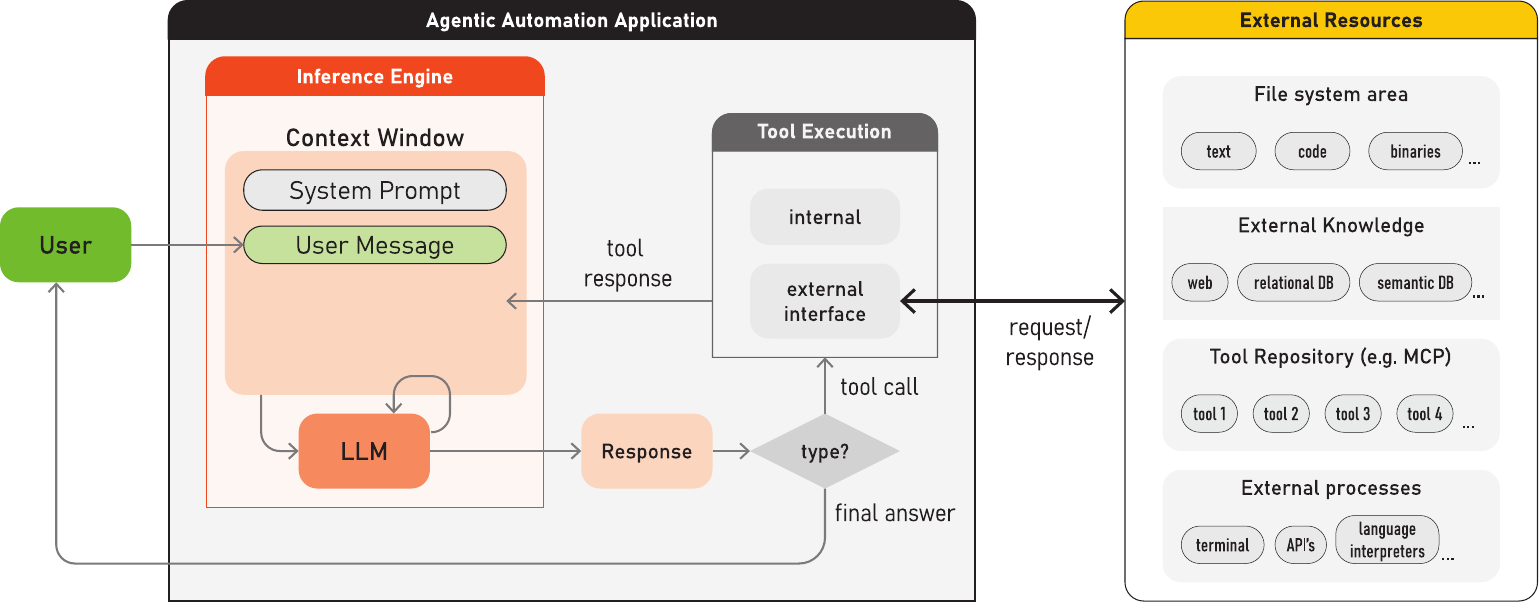}
    \caption{Architecture of an agentic application. The inference engine processes user messages through the LLM within a context window, generating either final answers or tool calls. Tool execution is handled internally or through external interfaces that connect to external resources including file systems, knowledge bases, tool repositories and system interfaces.}
    \label{fig:agentic-architecture}
\end{figure*}

This section addresses the gap identified in the previous sections by introducing the DUCTILE (Delegated, User-supervised Coordination of Tool- and document-Integrated LLM-Enabled) approach to agentic orchestration, whose core concept is to separate interpretation from computation. As argued in Sec. \ref{sec:backgroundsummary}, brittleness arises when adaptation is encoded in fixed rules. DUCTILE addresses this by assigning interpretation and orchestration to an LLM and delegating all calculations to verified tools, improving robustness to interface and input changes without relaxing engineering rigor. This separation also supports traceability and certification compliance, which are necessary for the aerospace context. It is aligned with the requirements R1--R12 summarized in Table \ref{tab:ductile-reqs}. 

The following subsections describe how the approach is implemented. The description is divided into three parts: developing the agentic implementation (Sec. \ref{subsec:development}), executing the agentic analysis (Sec. \ref{subsec:execution}) and evaluating the agentic performance (Sec. \ref{subsec:evaluation}). Fig. \ref{fig:agentic-architecture} shows the architecture of the agentic application used to execute the analysis.

\subsection{Developing the agentic implementation}
\label{subsec:development}

In the development stage, the agentic application is established so that orchestration can be observed, reproduced, and integrated with existing engineering tools. In line with the requirements (Table~\ref{tab:ductile-reqs}), the agentic application implementation is shaped by two considerations: (i) the \emph{character} of the agentic framework used to host the agent (addresses R1, R2, R9) and (ii) the \emph{configuration} of its core connections to models, prompts, and tools (addresses R2–R5, R8).

\subsubsection{Framework character}
The agentic framework exhibits inspectable and reproducible behavior, transparent configuration with minimal hidden state and low coupling to the analysis stack. In practice, this corresponds to open‑source, appropriately licensed solutions that surface prompts, traces, and file operations to the engineer, and that connect to external documents, tools, and analysis codes with minimal glue (addresses R1, R2, R9).

\subsubsection{Configuration scope}
Within such a framework, an agentic application is organized around three connective elements as shown in Fig. \ref{fig:agentic-architecture}: \emph{(a) the LLM model connection}, \emph{(b) the system prompt}, and \emph{(c) the tool interfaces}. The model connection covers the chosen model family and deployment mode (on‑premises or API), together with version pinning and environment capture so that runs are attributable across iterations (addresses R2, R5). The system prompt provides stable interaction conventions and output contracts, while domain procedures and method rules are provided \emph{at run time} through documents and skills rather than embedded as static instructions; this keeps behavior auditable and portable across projects (addresses R1, R9). Tool interfaces present deterministic, versioned capabilities to the agent via explicit, typed I/O contracts (paths, formats, units) and controlled write locations; generated scripts and invocation logs are persisted next to inputs/outputs to preserve traceability (addresses R3, R4, R9).

\subsubsection{Selecting and connecting to a LLM}
Model choice for engineering analysis is typically shaped by a three‑way trade‑off between capability profile, computational cost, and data‑governance constraints. Capability is commonly profiled along three dimensions: \emph{tool use} (selection and invocation of external tools), as reflected in $\tau$‑bench~\cite{barresTau2BenchEvaluatingConversational2025}; \emph{agentic reasoning} (multi‑step planning and recovery on error), as reflected in AgentBench~\cite{liuAgentBenchEvaluatingLLMs2025}; and \emph{code generation} quality, e.g., SWE‑bench~\cite{231006770SWEbenchCan}. The choice of on-premises vs externally hosted model deployment further conditions feasibility and governance. On‑premises execution offers full control over data flows, while externally hosted models reduce infrastructure burden and broaden immediate model availability. If an externally hosted solution is chosen, it shall be bounded by confidentiality requirements, export‑control rules, and applicable data‑protection frameworks (e.g., GDPR). For exploratory or non‑sensitive workloads however, external providers often function as a practical starting point. Regardless of the choice, model and version pinning enables run‑to‑run attribution (addresses R2, R5).

\subsubsection{System Prompt}
Model behavior is influenced by the system prompt \cite{zhangSPRIGImprovingLarge2024}, with design that can evolve iteratively or via systematic methods \cite{zhangSPRIGImprovingLarge2024,khattabDSPyCompilingDeclarative2023}. Guidance from major providers (OpenAI~\cite{GPT41PromptingGuide}, Anthropic~\cite{EffectiveContextEngineering}, Google~\cite{PromptDesignStrategies}, Meta~\cite{HowWriteEffective}, Mistral~\cite{PromptingMistralDocs}) offer different advice, but have common sections that are always recommended to include:
\begin{itemize}
    \item \textbf{Goal}: purpose and scope of responsibility.
    \item \textbf{Style}: tone and interaction conventions for communicating with the user.
    \item \textbf{Available tools}: typically injected automatically by the framework.
    \item \textbf{Reasoning strategy}: approach to task decomposition, depth of diagnosis and information exhaustiveness.
    \item \textbf{Output format}: a verbal schema or a structured contract (e.g., a final tool call).
\end{itemize}
Within the present approach, the prompt is kept task-agnostic (addresses R1, R9). Company‑specific procedures, methods, and design practices are not encoded statically; instead, they are retrieved at run time from knowledge resources or disclosed progressively via skills~\cite{Agentskills2026}. This keeps behavior auditable and portable, and allows the user to state the task (e.g., \emph{perform task X using method Y, revision Z}) while the agent assembles the relevant organizational information (addresses R1, R9, R12).

\subsubsection{Configuring tool access}
Tool access can be provided internally or via external resources (Fig.~\ref{fig:agentic-architecture}). The approach favors connecting to external resources wherever feasible to minimize framework complexity and maximize reusability across agent instances (addresses R9). Interfaces expose deterministic, versioned capabilities with strict I/O contracts (paths, formats, units) and controlled write locations; generated scripts and invocation logs (arguments and stdout/stderr) are persisted alongside inputs and outputs to preserve traceability and support inspection (addresses R3, R4).

\subsection{Executing the agentic analysis}
\label{subsec:execution}
Execution proceeds as an alternation between stochastic planning and deterministic tool use. The engineer provides the design context, inputs, and task objective; the application incorporates these into the context window. The inference process described in Sec.~\ref{sec:llm-foundations} proposes the next step toward the goal. Here, the stochasticity inherent to the model is an advantage: the model adapts inputs and company knowledge to the particular design context and goals (orange path in Fig.~\ref{fig:agentic-architecture}). Once the model proposes a tool call, the agentic framework executes it deterministically and appends the result to the context window. The agent repeats this cycle, calling successive tools until the task objective is met (addresses R3, R4, R7).

This loop connects the agent to the same tools an engineer would execute manually or hardcoded in a scripted pipeline, but the orchestration logic is no longer hardcoded. The LLM translates a \emph{natural language defined} process description from a document into a sequence of tool invocations, selecting each step based on the task objective, intermediate results, and available tools (addresses R1, R9). This adaptability is reported to the engineer in the final response and remains fully visible in the model traces and generated artifacts.

At any time in the parallel or sequential call of tools iterations, the agent may return the control to the user~\cite{yaoReActSynergizingReasoning2023}. The decision to finish the iteration is made by the LLM based on its training behaviors and the content of the context window. A few examples of when the LLM can choose (via the likelihood of the next token and subsequent generation of a response) to finish its turn are:

\begin{itemize}
    \item[-] When the request of the user can be answered directly, without the need to call tools.
    \item[-] When the model misses critical information to carry out the task.
    \item[-] When the tool calls have provided enough information to complete the task.
    \item[-] When a roadblock in the execution plan stops the model from getting closer to the goal (e.g., permission access, tool failure) and reasonable efforts have been made to circumvent those limitations within the independence scope defined in the context window.
\end{itemize}

\subsubsection{Final response and user intervention}
When stopping, the agent may summarize previous activities in the agent's turn. The agentic application parses the results and then finalizes the process by returning to the user the last message. Current agentic applications stream to the engineer directly the intermediate tool calls and thinking steps to follow the task progress. This allows for engineers to stop the agentic execution if it is deviating from the goal or stalling with an activity (addresses R6).

\subsubsection{Traceability and observability}
Messages, tool calls and tool results at minimum shall be sent for recording to an observability platform via standard payloads (e.g. OpenTelemetry). If a company external service is used, the same data sharing considerations shall be evaluated. Generated code is stored next to inputs/outputs to preserve traceability (addresses R3, R4). Where available, telemetry captures call latency and token/cost for model steps, supporting diagnosis, evaluation, and change control over time (addresses R8, R10, R11). The resulting traces provide the evidence required for do/check/approve review and for the evaluation described in the next sub-section.

\subsection{Evaluating the agentic application performance}
\label{subsec:evaluation}
Method adoption in engineering practice depends on evidence that the method behaves as expected, with a high degree of confidence \cite{seepersad2006validation}. Further, the system needs to be validated like any in-house engineering software with rigor proportional to the risk of the activity performed \cite{morrisonAssessingComputationalModel2019, raziRapidIntegrationNew2016}. However, agentic systems introduce challenges beyond those of traditional software.

First, the context window evolves across tool calls and intermediate reasoning steps, making the internal state difficult to inspect and debug. Observability platforms that trace each model invocation, tool call, and token exchange, such as Logfire~\cite{Logfire2024}, are essential for diagnosing failures (addresses R8). Second, outputs can vary structurally across identical inputs and yet can be equally valid. Because the DUCTILE approach intentionally uses flexibility to adapt to diverse engineering contexts, suppressing variability is not the goal. Instead, evaluation is performed against curated datasets of representative problems, including edge cases, similar to the held-out test sets used to assess generalization in machine learning \cite{hastieElementsStatisticalLearning2009}. These datasets verify that the agentic application produces correct and compliant outputs across the expected range of inputs, independent of the particular response path taken (addresses R10).

Unlike machine learning datasets containing only input $\mathbf{x}$ and expected output $\mathbf{y}$, each evaluation case must specify:
\begin{enumerate}
    \item The \emph{design environment}: the system prompt, the engineer query, the executable tools and reference documentation.
    \item The \emph{acceptance criteria}: expected outcomes expressed as quantitative tolerances, qualitative descriptions, or both.
    \item The \emph{evaluation method}: a deterministic check against reference values, a human expert review, or an LLM-as-a-judge assessment.
\end{enumerate}

The evaluation must cover both standard interactions and edge cases. Expected answers shall be developed and maintained by domain experts; in practice, these use cases grow to the order of hundreds of scenarios. Because model responses are probabilistic, each use case must be repeated to establish confidence in the agent's ability. DUCTILE adopts the pass\textasciicircum{}k metric~\cite{yaoTauBenchBenchmarkToolAgentUser2025}, defined as whether the agent correctly performs the same task in $k$ independent repetitions. The choice of $k$ depends on the evaluation scenario and the criticality of the task.

A useful analogy comes from material qualification in aerospace structures, where allowable properties are defined probabilistically as $x/y$: the $x$-th population percentile at $y$\% confidence. Three standard levels exist: S, B, and A-Basis~\cite{MMPDS}. Treating each evaluation case as an independent Bernoulli pass/fail event, the Clopper--Pearson exact bound~\cite{clopperUseConfidenceFiducial1934} for $k$ successes in $k$ trials yields the minimum number of consecutive passes required at each confidence level, as shown in Table~\ref{tab:pass-k}.

\begin{table}[ht]
    \caption{Minimum consecutive passes $k$ for the pass\textasciicircum{}k metric at 95\% confidence, derived from the Clopper--Pearson bound $p_{\mathrm{lower}} = \alpha^{1/k}$ with $\alpha = 0.05$.}\label{tab:pass-k}
    \centering
    \begin{tabular}{lccc}
        \toprule
        & S-Basis & B-Basis & A-Basis \\
        \midrule
        Pass probability $p$ & $\geq 0.50$ & $\geq 0.90$ & $\geq 0.99$ \\
        Required $k$         & 5           & 29          & 299         \\
        \midrule
        \multicolumn{4}{l}{\emph{DUCTILE recommended values}} \\
        Development           & \multicolumn{3}{c}{$k = 3$} \\
        Deployment            & \multicolumn{3}{c}{$k \geq 10$} \\
        \bottomrule
    \end{tabular}
\end{table}

The basis-level $k$ values serve as a reference for comparison with established engineering practice. In practice, DUCTILE delegates execution but not responsibility. The engineer retains accountability for reviewing outputs, and existing engineering quality processes remain in place. Since each repetition consumes computational resources and cost, the DUCTILE approach recommends $k = 3$ during development and $k \geq 10$ for deployment, increasing $k$ as the criticality or autonomy of the agent grows.

Evaluations must be re-run whenever the system prompt, tools, documentation, or underlying model changes. At the scale of hundreds of cases repeated $k$ times, human evaluation becomes unfeasible. Automated assessment via deterministic checks or LLM-as-a-judge is therefore essential, with trust metrics analogous to inter-rater reliability established for the latter.

%% file: 05-case-example.tex
\section{Industrial application of agentic load processing on a realistic case}
\label{sec:case-example}

\begin{figure*}[hbt!]
    \centering
    \includegraphics[width=\textwidth]{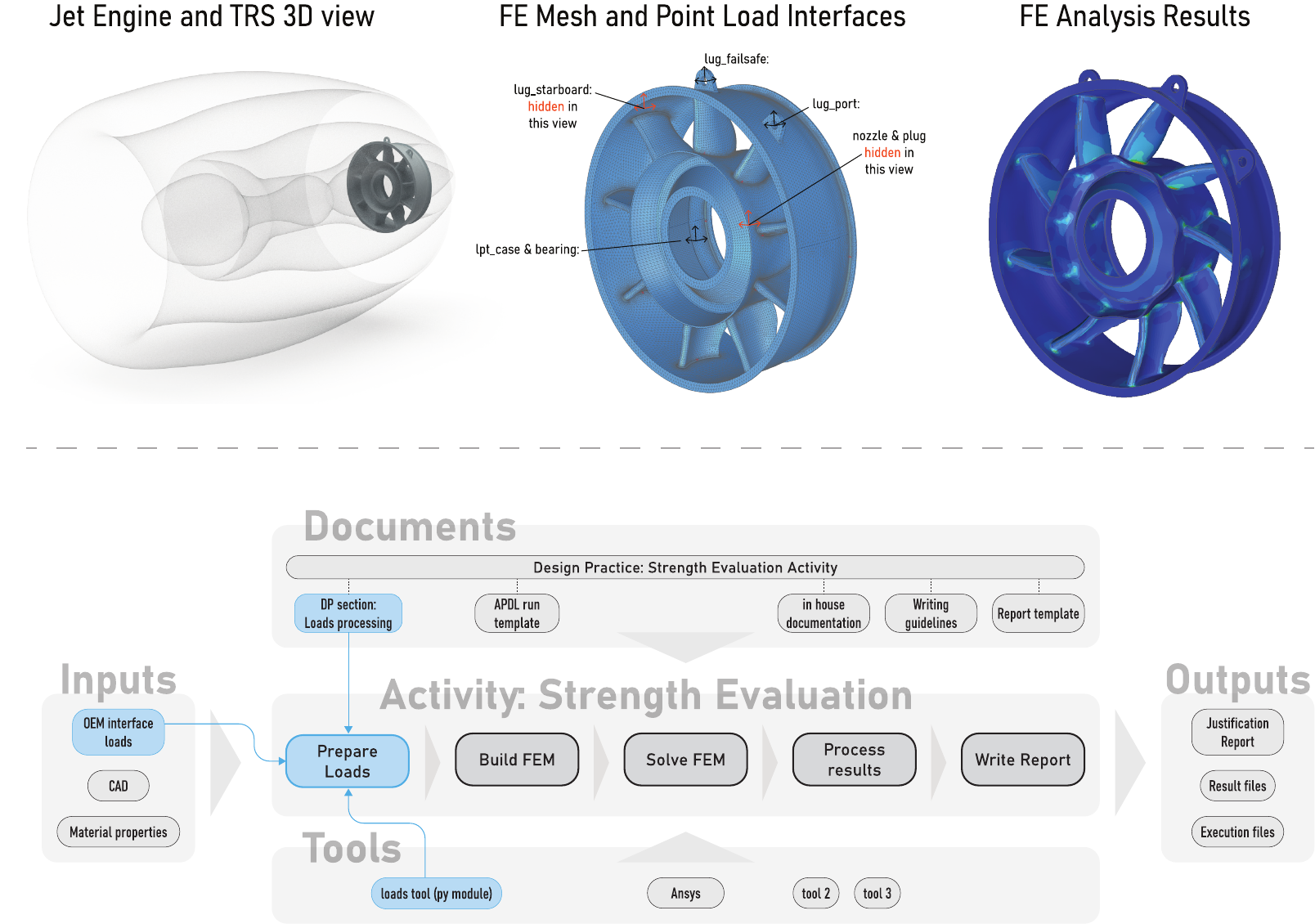}
    \caption{Top: Views of the component in the physical domain. Bottom: Activity View of the TRS Strength evaluation. Only a subset of the inputs, documents and tools are shown for visual clarity. Highlighted in orange is the load processing step where the use case focuses.}
    \label{fig:trs-geometry}
\end{figure*}

The DUCTILE approach described in Sec. \ref{sec:DUCTILE} is applied to a structural strength evaluation on a realistic case at the case company. The component in this case is a Turbine Rear Structure (TRS), shown in Fig. \ref{fig:trs-geometry}. While the case company routinely performs static strength analyses on the TRS as part of the certification and continuous product development activities, the jet engine OEM generally owns the global stiffness and loads models at the whole-engine integration level, and delegates structural compliance checks to component partners. A design practice document at the case company defines the steps, methods, and expected outputs for this type of analysis.

\textbf{The use case description.} The engineering use case concerns a new structural evaluation triggered by a design change in a neighboring engine component. The geometry, finite element mesh and analysis methods remain unchanged; only the loading input from the OEM is new. A new, company-maintained Python-based loads processing tool is used together with the established design practice.

The new OEM delivery introduces four deviations compared to the previous analysis. Three of these deviations are input format changes, a common example of variability when data crosses organizational boundaries; the fourth is a correction method communicated by the OEM requiring a 1.04 factor on all Fx forces, separate from the raw data. In a traditional scripted pipeline, each deviation would require manual intervention or code modification. Table \ref{tab:deviations} summarizes the four deviations.

\begin{table}[ht]
    \caption{Deviations in the new analysis scenario. The first three are input format changes; the fourth is an OEM-communicated correction specified in the task description.}\label{tab:deviations}
    \centering\small
    \begin{tabularx}{\columnwidth}{@{}p{1.6cm}XX@{}}
        \toprule
        Deviation & What changed & Impact on automation \\
        \midrule
        File format & YAML instead of JSON & Requires new parser \\[3pt]
        Unit system & Imperial instead of SI & Silent error  \\[3pt]
        Node naming & \texttt{right/left} instead of \texttt{port/starboard} & FEM run fails downstream \\[3pt]
        Method change & 1.04 factor on Fx forces  & Silent error \\
        \bottomrule
    \end{tabularx}
\end{table}

The remainder of this section describes the configuration as shown in Sec. \ref{subsec:case-setup}, the execution with two engineers (Sec. \ref{subsec:case-execution}) and the systematic evaluation of the agent's performance, in Sec.~\ref{subsec:case-evaluation}.

\subsection{Case Setup}
\label{subsec:case-setup}

Following Sec. \ref{subsec:development}, the agentic implementation was built using Claude Code \cite{ClaudeCode2025}, a terminal-based agentic framework. Extensive description of the implementation is available in the supplemental material. Claude Code operates within the engineer's existing IDE and file system. Since the engineers at the case company already work with scripts, input files, and terminal commands, the agentic application fits within the existing workflow rather than imposing a new one. The engineer can inspect every file the agent reads or writes (addresses R1, R2). To maximize transparency and user experience, engineering product data information was placed in the working file system during the experiments. The design practice was served via an MCP document server, giving the agent version controlled access to the methodology. The loads processing tool was distributed as a Python package via PyPI, with API documentation published in both human-readable and LLM-consumable formats (\texttt{llms.txt}) on the corresponding GitHub Pages.

The task description was prepared prior to agent testing, targeting an experienced engineer or an inexperienced graduate joining the company; no refinements to the prompt or integration were required after initial testing. The design practice document captures the logic of the activities to be performed. Fig. \ref{fig:loads-flowchart} shows the workflow derived by the authors from this document. This flowchart was not available to the agent. By reading the document and inspecting the inputs, the agent was expected to infer and implement a comparable sequence of activities (addresses R1, R12).

\begin{figure}[tb]
\centering
\begin{tikzpicture}[
  node distance=0.20cm and 1.4cm,
  >={Stealth[length=5pt, width=3.5pt]},
  process/.style={
    rectangle, rounded corners=2pt, draw=chalmersLeaves200, fill=white,
    text=black, font=\small, text width=4.8cm, minimum height=0.5cm,
    align=center, inner sep=2pt},
  decision/.style={
    diamond, draw=chalmersLeaves200, fill=chalmersLeaves50, text=black!80,
    font=\small, aspect=3, inner sep=1pt, align=center},
  startstop/.style={
    rectangle, rounded corners=5pt, draw=chalmersLeaves700, fill=chalmersLeaves700,
    text=white, font=\small\bfseries, minimum width=1.6cm,
    minimum height=0.45cm, align=center},
  yeslabel/.style={font=\small\bfseries, text=chalmersLeaves700},
  nolabel/.style={font=\small\bfseries, text=chalmersLeaves700},
  every path/.append style={draw=chalmersLeaves200},
]

\node[startstop]                        (start)  {Start};
\node[process, below=of start]          (read)   {Read OEM load file \& verify format};
\node[process, below=of read]           (down)   {Downselect applicable load cases};
\node[decision, below=0.38cm of down]   (sicheck){OEM delivery\\in SI units?};
\node[process, below=0.38cm of sicheck] (units)  {Convert units to FEM system};
\node[process, below=of units]          (equil)  {Verify equilibrium for each load case};
\node[decision, below=0.38cm of equil]  (ult)    {Ultimate analysis\\with limit loads?};
\node[process, below=0.38cm of ult]     (factor) {Apply 1.5 ultimate load factor};
\node[process, below=of factor]         (write)  {Write one \texttt{.inp} file per load case};
\node[process, below=of write]          (envtab) {Produce summary envelope table};
\node[decision, below=0.38cm of envtab] (prev)   {Previous loads\\available?};
\node[process, below=0.38cm of prev]    (exceed) {Produce exceedance comparison};
\node[startstop, below=of exceed]       (stop)   {End};

\draw[->] (start)  -- (read);
\draw[->] (read)   -- (down);
\draw[->] (down)   -- (sicheck);
\draw[->] (sicheck)-- node[right, nolabel]{No} (units);
\draw[->] (units)  -- (equil);
\draw[->] (equil)  -- (ult);
\draw[->] (ult)    -- node[right, yeslabel]{Yes} (factor);
\draw[->] (factor) -- (write);
\draw[->] (write)  -- (envtab);
\draw[->] (envtab) -- (prev);
\draw[->] (prev)   -- node[right, yeslabel]{Yes} (exceed);
\draw[->] (exceed) -- (stop);

\draw[->] (sicheck.east) -- ++(1.2,0)
  node[above, yeslabel, pos=0.3]{Yes}
  |- (equil.east);

\draw[->] (ult.east) -- ++(1.2,0)
  node[above, nolabel, pos=0.3]{No}
  |- (write.east);

\draw[->] (prev.east) -- ++(1.2,0)
  node[above, nolabel, pos=0.3]{No}
  |- (stop.east);

\end{tikzpicture}
\caption{Engineer-interpreted activity process flowchart of the natural language and unstructured description contained in the loads design practice document. NOTE: The agent must infer a similar workflow by reading the document; this flowchart is not available in the agent's context.}
\label{fig:loads-flowchart}
\end{figure}

\subsection{Execution}
\label{subsec:case-execution}

Two engineers at the case company were assigned the same task with the same inputs and deviations described in Table \ref{tab:deviations}. Both had access to the agent and were encouraged to interact with it and verify its outputs as they saw fit. The transcripts and output files are available in the supplemental material.

\subsubsection{Engineer 1: Delegation}

Engineer 1, familiar with agentic workflows, delegated the task to the agent and checked the outputs after the full activity was completed. Fig. \ref{fig:engineer1-chat} summarizes the interaction.

The engineer pointed the agent to the task description and requested a plan. The agent read the task description PDF, fetched the design practice via the MCP document server, explored the working directory, and examined the previous analysis run and input files. During this reading phase, the agent identified all four deviations in its reasoning trace: the YAML format incompatible with the tool's JSON parser, the imperial unit system, the changed pilot node names, and the 1.04 correction factor specified in the task description. The agent surfaced these findings in a seven-step plan presented to the engineer, proposing a processing approach that addressed each deviation. The engineer approved the plan.

The agent then fetched the tool's API documentation, wrote a processing script, executed it and verified the outputs: six \texttt{.inp} files (one per envelope-selected load case), an envelope summary table, and an exceedance comparison against the previous analysis. The engineer inspected the generated script and output files independently and declared the task completed.

\begin{figure}[!p]
\rightactionbubble{chalmersLeaves25}{chalmersLeaves200}{%
  {\sffamily\bfseries\scriptsize Engineer 1}\par\smallskip
  \small\itshape Opens the input files and design practice in the IDE. Reviews the deviations independently.%
}
\vspace{1pt}

\rightbubble{chalmersLeaves10}{chalmersLeaves200}{%
  {\sffamily\bfseries\scriptsize Engineer 1}\par\smallskip
  \small plan this: @ref\_documents/task\_description.pdf%
}
\vspace{1pt}

\leftbubble{chalmersBrick10}{chalmersBrick400}{%
  {\sffamily\bfseries\scriptsize Agent}\par\smallskip
  \toolcall{read(task\_description.pdf)}\par
  \toolcall{bash(ls) working directory}\par
  \toolcall{get\_document\_content(MCP) design practice}\par
  \toolcall{read(previous\_run/\ldots)}\par
  \toolcall{read(OEM\_loads\_v2.yaml)}\par\smallskip
  \thinkbubble{The new file is YAML but the tool expects JSON\ldots\ Units are in imperial, the design practice requires SI\ldots\ Pilot nodes are named \texttt{left/right} instead of \texttt{port/starboard}\ldots\ The task description specifies a 1.04 Fx correction factor not present in the previous run\ldots}%
}
\vspace{1pt}

\leftbubble{chalmersBrick10}{chalmersBrick400}{%
  {\sffamily\bfseries\scriptsize Agent}\par\smallskip
  \small Here's the plan:\par\smallskip
  \planbubble{%
    \hangindent=1.5em\hangafter=1\noindent 1.\ Read YAML, rename \texttt{lug\_left}$\rightarrow$\texttt{lug\_port}, \texttt{lug\_right}$\rightarrow$\texttt{lug\_starboard}.\par
    \hangindent=1.5em\hangafter=1\noindent 2.\ Apply OEM Fx correction ($\times$1.04).\par
    \hangindent=1.5em\hangafter=1\noindent 3.\ Verify coordinate system (no transform).\par
    \hangindent=1.5em\hangafter=1\noindent 4.\ Convert units klbs $\rightarrow$ N.\par
    \hangindent=1.5em\hangafter=1\noindent 5.\ Envelope downselection.\par
    \hangindent=1.5em\hangafter=1\noindent 6.\ Write \texttt{.inp} files (exclude bearing).\par
    \hangindent=1.5em\hangafter=1\noindent 7.\ Envelope table + exceedance comparison.%
  }%
}
\vspace{1pt}

\rightbubble{chalmersLeaves10}{chalmersLeaves200}{%
  {\sffamily\bfseries\scriptsize Engineer 1}\par\smallskip
  \small yes%
}
\vspace{1pt}

\leftbubble{chalmersBrick10}{chalmersBrick400}{%
  {\sffamily\bfseries\scriptsize Agent}\par\smallskip
  \toolcall{web\_fetch(ductile-loads/llms-full.txt)}\par
  \toolcall{write(loads\_processing\_v2.py)}\par
  \toolcall{bash(uv run loads\_processing\_v2.py)}\par\smallskip
  \small Verifying outputs\ldots\par
  \toolcall{bash(ls limit\_loads/)}\par
  \toolcall{read(limit\_load\_2.inp)}\par
  \toolcall{read(comparison\_report/v1\_vs\_v2.json)}\par\smallskip
  \small Processing complete. 6 load case files generated. New loads exceed previous envelope --- Mz~+30\%, Fx~+15\%, Fy~+11--17\%.%
}
\vspace{1pt}

\rightactionbubble{chalmersLeaves25}{chalmersLeaves200}{%
  {\sffamily\bfseries\scriptsize Engineer 1}\par\smallskip
  \small\itshape Checks the generated script and output files. Declares the task completed.%
}
\caption{Summary of Engineer 1's interaction. Green bubbles: engineer messages. Orange bubbles: agent responses. Dashed boxes: agent reasoning. Yellow boxes: tool calls. Dotted boxes: engineer actions outside the conversation. An expanded transcript is available in the supplemental material.}
\label{fig:engineer1-chat}
\end{figure}

\subsubsection{Engineer 2: Incremental discovery}

A second engineer, unfamiliar with agentic workflows, took the opposite approach. Rather than delegating the full task, this engineer requested one step at a time: first reading the inputs, then processing a single load case, then checking the output before proceeding. Each intermediate result was inspected before the next step was requested. Through this incremental process, the engineer discovered the task, learned the tool's behavior, and built confidence in the agent gradually. The agent adapted to this step-by-step supervision without difficulty. The final output was identical to Engineer 1's result: the same load case files, envelope table and exceedance comparison. The same four deviations were handled, through a different supervision path.

\subsection{Evaluation}
\label{subsec:case-evaluation}

Following the evaluation approach described in Sec. \ref{subsec:evaluation}, the agent's performance was assessed against the scenario defined by the four deviations in Table \ref{tab:deviations}. After an independent check by a company domain expert, Engineer 1 output data was used as ground truth for automated evaluation.

The evaluation used two complementary checks. First, an LLM-as-a-judge assessment (Opus 4.6) verified that the agent's processing script correctly addressed all four deviations: YAML-to-JSON conversion, unit conversion, pilot node mapping, and application of the 1.04 Fx correction factor. Second, a deterministic comparison checked the final load summary values against the expert's reference output. A run passed only if both checks succeeded. The LLM-as-a-judge assessments were manually reviewed by the domain expert, yielding 100\% inter-rater agreement.

The agent was run 10 independent times on the same scenario, with no modifications to the prompt, tools, or context between runs. All 10 runs passed both quantitative and qualitative checks, yielding a 100 per cent inter rater reliability by LLM-as-a-judge (Opus 4.6) and the 10 manual verifications. The evaluation scripts and outputs are available in the supplemental material. The implementation is an example of the \mbox{DUCTILE} approach towards agentic evaluation.

%% file: 06-discussion.tex
\section{Discussion}
\label{sec:discussion}

\subsection{Closing the Brittleness Loop}
\label{subsec:brittleness-loop}

Sec.~\ref{sec:introduction} identified brittleness as a structural property of engineering analysis automation, breaking rigid pipelines when changes occur. In the case study of Sec.~\ref{sec:case-example}, the agentic application consistently identified the errors that would have broken the downstream processes, and adapted to changes consistently with pass\textasciicircum{}k=10. Therefore it was demonstrated using the DUCTILE agentic orchestration approach how to develop, use and evaluate an agentic application that meets the engineering analysis performance expectations and fulfills the requirements identified in Table~\ref{tab:ductile-reqs}. It was achieved without the need to maintain product datasets, or company specific LLMs. While it was exemplified on a pre-processing task for a mechanical activity, it is left to the reader to interpret the same orchestration and support in a post-processing task, or for other engineering analyses where in-house tools are combined with external tools and written procedures.

\subsection{Consequences for Practice}
\label{subsec:consequences}

The application of agentic support exposed the gaps in the existing design practice documentation at the case company. Experienced engineers bridge ambiguities in written procedures through tacit knowledge. Hsieh et al.~\cite{hsiehToolDocumentationEnables2023} demonstrate that well-written tool documentation enables LLMs to use unfamiliar tools zero-shot, without fine-tuning. Yuan et al.~\cite{yuanEASYTOOLEnhancingLLMbased2024} show that concise, structured descriptions outperform verbose ones. Investing in documentation quality serves both human engineers and agents. However, the development of the use case showed how those tacit activities were missing from the company design practices, and had to be added while the agentic application development and evaluations were being co-developed. Moreover, by requiring engineering knowledge to be captured in explicit, accessible documents rather than residing in individual expertise alone, the use case showed how the approach reduces the organizational vulnerability that arises when critical knowledge is concentrated in individuals.

The DUCTILE agentic approach uses in-context reasoning exclusively (no fine-tuning nor persistent memory). This is a deliberate choice because, together with evaluations, it allows the agentic application to avoid model lock-ins, and to update the model as smaller-footprint, cheaper, or more capable models are released without effort and robustly.
This contrasts with post-training approaches where tool-use strategies are embedded in model weights~\cite{weiAgenticReasoningLarge2026}, and with rule-based systems that, despite formally explicit logic, become practical black boxes as rule chain complexity exceeds what practitioners can trace (Sec.~\ref{sec:design-automation}). Both engineers in the case study confirmed that the agent's actions were easy to follow and inspect, showing promise for wider adoption in practice with minimal maintenance overburden.

Despite the case for a dataset-free configuration pipeline, more complex scenarios may call for company- or task-specific models: cases where agentic performance (tested via evaluations) is not robust enough despite optimizing the context window, or where latency must be minimized and context length reduced. In such scenarios, the parameters of an open-weights model can be fine-tuned to inject company knowledge, and this is where a training dataset becomes necessary. Datasets in engineering, however, remain scarce~\cite{ahmedDesignByData2025}. Ironically, adopting the agentic approach itself generates the raw material. Every engineer-agent session automatically produces a conversation log capturing design intent and rationale, even when the agent performs poorly. This history can be passed to a separate LLM pipeline to generate positive examples of expected behavior, effectively producing a training dataset without dedicated engineering effort.

\subsection{Risks and Unintended Consequences}
\label{subsec:risks}

The aerospace community warns that ``overreliance on software tools can lead to a superficial understanding of the underlying physics''~\cite{KrupaArtScienceCreative}. This concern applies to agents as well. A key distinction is needed, however, between dexterity and judgment. Dexterity with specific APIs, data formats, and tool interfaces is what the agent will absorb. Engineering judgment, knowing whether results make physical sense, the method is appropriate, or the output meets certification requirements, must remain with the engineer. In our experience developing aerospace products under intense project deadlines, the real risk is in the organizations stopping to recognize and allocate resources to develop the engineer's judgment. Engineer~2's step-by-step interaction pattern provides evidence that supervision and learning can coexist: this engineer built understanding of the task progressively while the agent handled the implementation details.

Gericke and Eckert~\cite{gerickeCoevolutionToolsMethods2026} raise two cautions from the product development ecosystem perspective. First, a rebound effect: historically, tools that shortened development time led to more complex products requiring similar development time. Second, skill loss: designers may lose tacit skills previously developed through routine manual execution. Therefore, we call for a conscious implementation. The supervisory role must be deliberately designed to maintain engineering judgment~\cite{bainbridgeIroniesAutomation1983}.  

It remains to be seen whether agentic support will enrich or exhaust the work of engineers. On one hand, many design and AI researchers, including Cross~\cite{cross2025natural}, argue that these systems that interactively support humans should be designed in a way that is \emph{cognitively comfortable} for them. If implemented in a human-centered way~\cite{liWorldsISee2023}, the mundane tasks could be absorbed by the agent, and the most rewarding aspects of engineering work preserved~\cite{liBEHAVIOR1KBenchmark2023}.  However, we are starting to observe in software engineering how these agents are not delivering this promise. Instead, the joy of solving the little problems and the social nature of the work are reduced, being substituted by a hollow supervision role that exhausts software engineers via increasing the intensity of their work~\cite{summersHumanInLoopTired2026}. Shneiderman~\cite{shneidermanHumanCenteredAI2022} argues that this outcome is not inevitable: systems designed as ``supertools'' that amplify human capability, rather than autonomous agents requiring supervision, can preserve both productivity and satisfaction. The choice of implementation, however, belongs to the engineering organizations. Ultimately, they must resist the temptation to optimize solely for short-term efficiency, and instead design an automation support that enhances productivity while improving the quality of work and developing engineering judgement.

\subsection{An Engineering Automation Paradigm Shift}
\label{subsec:paradigm-shift}

Engineering automation has historically required deterministic logic, but LLM-based agents violate this expectation. They are probabilistic, their reasoning is not formally verifiable, and they can produce incorrect outputs. Based on our previous work~\cite{gomezTeamThreeRole2025} and interactions at the case company, we expect their acceptance for certification activities to be contested.

This contest echoes a recurring pattern in AI where hand-crafted, ontology-driven methods are eventually displaced by statistical ones once performance is demonstrated at scale. Von Neumann~\cite{neumannProbabilisticLogicsSynthesis1956} showed in 1956 that reliable systems can be built from unreliable components through redundancy. Rosenblatt himself~\cite{rosenblattPerceptron1957} acknowledged the limitations of probabilistic methods: \emph{the penalty that we pay for the use of statistical principles in the design of the system is a probability that we may get a wrong response in any particular case}. Nevertheless, Minsky and Papert~\cite{minskyPerceptrons1969} dismissed neural networks in 1969. It was not until sufficient computational power and dataset size became available that neural networks prevailed, as demonstrated by AlexNet~\cite{krizhevskyImageNetClassificationDeep2012}. Sutton~\cite{suttonBitterLesson2019} generalized this observation: methods that leverage computation consistently outperform methods that leverage human knowledge. In this framing, the probabilistic approach that produces occasional errors and the flexibility that absorbs the design process variability are two sides of the same coin. One cannot exist without the other. We argue for the acceptance of such probabilistic agentic orchestration to be judged by the merit of the task evaluations (Sec.~\ref{subsec:evaluation}), and not against pre-conceived deterministic expectations.

\subsection{Limitations}
\label{subsec:limitations}

Several limitations constrain the generalizability of these findings.

\begin{itemize}
    \item \textbf{Two engineers.} The experiment is illustrative, two participants' behaviours cannot be extrapolated to the engineering population.
    \item \textbf{Model dependence.} The results reflect Claude Opus~4.6 at the time of the study and may change with future versions and other models.
    \item \textbf{Requires documented processes.} Where engineering knowledge is predominantly tacit and not captured in accessible documents and tools, the agent has nothing to read and the approach does not apply.
    \item \textbf{Text and data workflows only.} This work does not cover spatial tasks such as CAD geometry generation, finite element meshing, or graphical post-processing. Image modality interpretation has not been explored.
    \item \textbf{Transferability not demonstrated.} Applicability to other engineering domains, organizations and task types is argued on structural grounds but not empirically validated beyond this case.
\end{itemize}

\subsection{Future Work}
\label{subsec:future-work}

\emph{An open dataset for agentic evaluation is necessary}. Organizations deploying agents in certified engineering need references for developing internal datasets that capture their specific tools, design practices and quality requirements. Any change to the underlying model, prompts, tool version or design practice can alter agent behavior and must be assessed before deployment. The research community, in turn, needs open engineering benchmarks. Efforts such as calls for curated design datasets~\cite{ahmedDesignByData2025} and EngiBench~\cite{feltenEngiBench2025} have begun to address it for generative tasks, and similar calls have emerged for LLM evaluation in engineering~\cite{mustaphaLLMsMechanicsDesignManufacturing2025}. Agentic engineering evaluations have additional, nuanced requirements: the framework must be executable and available to simultaneously apply domain knowledge, call tools correctly, interpret numerical outputs and follow documented methods.

\emph{Industrial impact shall be measured}. The two engineers in Sec.~\ref{sec:case-example} interacted with the agent in fundamentally different ways. One delegating the full task, the other proceeding incrementally. A controlled within-subject study comparing engineer performance with and without agent support would formalize this observation and measure impact. Metrics like task accuracy, completion time and confidence in the result across a larger sample of practicing engineers would reveal how engineering practices interact in realistic scenarios.

%% file: 07-conclusion.tex
\section{Conclusion}
\label{sec:conclusion}

The presented DUCTILE approach demonstrated the automation support separation via an industrial load processing task at an aerospace manufacturer. The adaptive orchestration was performed by the agent, using the non-deterministic nature of the LLMs. The deterministic execution was performed by verified engineering tools. Engineers retain responsibility for supervision and judgment while the process is guided by design practices already available in companies.

In the use case, the agent handled four input deviations that would break traditional scripted automation. All four were resolved correctly consistently across 10 independent evaluation runs. The approach was also tested on two engineers with different supervision styles, one delegating the full task and one proceeding incrementally, and both reached the same correct result using the same agent configuration.

Traditional engineering automation is brittle because it encodes adaptation in deterministic rules bound to interfaces that, in practice, keep changing. When the ecosystem changes, when people move, the automation support becomes a barrier and stops being an enabler. The DUCTILE agentic approach presented here absorbs that variability at the orchestration layer, so that engineering time and cognitive effort are spent on the decisions that affect product performance rather than on the data wrangling and workflow adaptation that currently consume it.